\documentclass[referee]{aa}
\usepackage{graphicx}
\begin{document}
\title{2D MHD modelling of compressible and heated coronal loops obtained via
nonlinear separation of variables and compared to TRACE and SoHO observations}

\titlerunning{MHD Loop Models}
\authorrunning{Petrie et al.}

\author{\bf G.J.D.~Petrie
          \inst{1}
          \,
          C. Gontikakis
          \inst{2}
          \,
          H. Dara 
          \inst{2}
          \,
          K. Tsinganos
          \inst{1}
          \and
          M.J. Aschwanden
          \inst{3}
          }

   \offprints{G.J.D. Petrie}

   \institute{IASA and Section of Astrophysics, Astronomy and Mechanics,
Department
of Physics, University of Athens, Panepistimiopolis, GR-157 84 Zografos,
Athens, Greece
              \email{gordonp@phys.uoa.gr,tsingan@phys.uoa.gr}
         \and
              Academy of Athens, Research Center of Astronomy and Mathematics,
GR-106 73 Athens, Greece\\
              \email{cgontik@cc.uoa.gr,edara@cc.uoa.gr}
         \and
             Lockheed Martin, Solar and Astrophysics Lab. L9-41, Bldg. 252,
3251 Hanover Street, Palo Alto CA 94304\\
              \email{aschwanden@lmsal.com}
             }

   \date{Received;accepted}

\abstract{
An analytical MHD model of coronal loops with compressible flows and 
including heating is compared to observational data.  
The model is constructed via a systematic nonlinear separation of 
the variables method used to calculate several classes of exact MHD 
equilibria in Cartesian geometry and uniform gravity. 
By choosing a particularly versatile solution class with a large 
parameter space we are able to calculate models whose loop length, 
shape, plasma density, temperature and velocity profiles are fitted 
to loops observed with TRACE, SoHO/CDS and SoHO/SUMER.  Synthetic emission
profiles are also calculated and fitted to the observed emission patterns. 
An analytical discussion is given of the two-dimenional balance of the 
Lorentz force and the gas pressure gradient, gravity and inertial 
forces acting along and across the loop. 
These models are the first to include a fully consistent description 
of the magnetic field, 2D geometry, plasma density and temperature, 
flow velocity and thermodynamics of loops.  
The consistently calculated heating profiles which are largely dominated 
by radiative losses are influenced by the flow and are asymmetrical 
being concentrated at the upflow footpoint. 
}

   \maketitle
%

\section{Introduction}
\label{introduction}
A significant proportion of the energy emission from the solar
corona is concentrated along loops which are believed to trace closed lines of
force of the
magnetic field, which penetrates the photosphere from below and
expands to fill the whole of the coronal volume above an active
region (Bray et al.,~1991). A coronal loop is therefore an
important localised
structure which connects the photosphere to the corona through 
the transition region and may thus be studied to gain information 
about the heating of the corona as a whole (Aschwanden,~2003).

Early results from the Skylab mission emphasising that the solar corona is not
a
homogeneous medium but filled with loop structures stimulated much interest in
modelling those loop structures.  The first models were one-dimensional
hydrostatic models which balanced heat conduction and radiative losses with an
imposed heating function.  Rosner et al.~(1978) balanced radiative losses and
heat conduction
against heating assuming zero heat conduction across the foot points, a
restriction relaxed by Hood \& Priest~(1979) while they neglected radiative
losses.  Vesecky et al.~(1979), Serio et al.~(1981) and Wragg \& Priest~(1981, 
1982)
added the effects of varying pressure and gravity to their models, as well as 
the effects of a variable loop cross-sectional area. 
Cargill \& Priest~(1980) were the first to add adiabatic plasma flows and 
concentrated on examining the relationship between cross-sectional area and 
flow velocity along the loops, while later (1982) introduced non-adiabatic
flows 
balancing the net heat in/out against conduction and radiation with a heating 
function proportional to the density.  Further important hydrodynamic modelling
of plasma flows in solar
atmospheric structures has been carried out and applied to photospheric flux
tubes by Thomas and others, work summarised in Thomas~(1996).

A similarly strong interest in loop modelling in recent years has been
motivated 
by the higher resolution results from the Yohkoh, SoHO and TRACE spacecrafts.  
A systematic study of a one-dimensional hydrodynamic solution class of loops 
with constant cross-section has been carried out by Orlando et
al.~(1995a,b), with non-adiabatic flows and balancing the net heat in/out
against conduction, radiation and a heating function.  Much attention has
focused on the form of the heating function as a means of inferring the 
coronal 
heating mechanism.  
Priest et al.~(1998, 2000) concluded that their loop model fitted to Yohkoh
data 
yielded higher heating at the apex, although re-analysis of this Yohkoh loop 
by
Mackay et al.~(2000) and Aschwanden~(2001)
concluded that the loop was heated at the foot points.  Aschwanden et
al.~(2001) systematically explored a class of one-dimensional hydrostatic
solutions with a non-uniform heating function in exponential form and fitted
them to a large sample of EUV loops observed with TRACE.  They found that most
of the sample of loops could not be modelled by their hydrostatic solution
class, and that those which could were heated near the foot points.  
In the present paper we use a different approach.  We do not impose a priori 
a specific form for the heating function, but instead we calculate it after 
completing a fitted dynamical MHD model.  Hence we calculate a model for 
the observable quantities first and then find a consistent heating function 
from the first law of thermodynamics.

Of course some loops in the solar atmosphere are far from
equilibrium and time-dependent hydrodynamical loop models have also been
developed by e.g. Mariska \& Boris~(1983), Cargill~(1994), Cargill \&
Klimchuk~(1997), Walsh et al.~(1995, 1996), Walsh \& Galtier~(2000),
Peres~(2000), Reale et al.~(2000a, b).  In this paper we restrict our
modelling and observations to steady-state loops.

To date all heated loop studies have included only one-dimensional hydrostatic
or hydrodynamic models.  However, in the highly magnetised and sparse coronal
plasma the magnetic field is likely to have a significant direct effect on the
statics or dynamics of such a curved structure as a coronal loop, while plasma
flow, even at such sub-Alfv\'enic velocities as $20$~km/s (Dara et al., 2002),
may have an effect on the heating balance.  Moreover, the geometrical
details may have an impact on the energy profile of the loop, via the
potential energy, and therefore on the heating model.  The models in this paper
include
two-dimensional geometry, compressible MHD plasma flow in uniform
gravity and heating in single consistent exact solutions and thereby give a
first opportunity to investigate these
effects.  We fit the geometrical and dynamical aspects of the
model to a loop observed by TRACE, a case observed with SoHO/CDS
(Schmelz et al.~,2001) and a loop observed by
SoHO/SUMER. We also 
give a model of the energy balance of each loop, including the 
loop heating.

The paper is organised as follows.  The solution class is described in Sect.
\ref{thesolutionclass}, and the method of constructing the models is explained
in
Sect. \ref{construction}.  The observations and data analysis is described in
Sect. \ref{observations} and models fitted to data sets are presented in Sect.
\ref{models}.  The paper is concluded with Sect. \ref{conclusions}.

\section{The analytical model}
\label{analyticalmodel}
In this section, after an introduction of the basic equations needed in order
to 
establish notation, we proceed to a brief presentation of the key assumptions 
for the derivation of the particular solution class and an outline of the
method 
employed for the construction of the particular solutions.   
\subsection{Basic equations}
\label{basicequations}

Our model utilize solutions obtained by using a systematic nonlinear
separation of the variables construction method in two dimensions and 
Cartesian geometry (Petrie et al.~ 2002, henceforth, Paper 1), already seen 
in spherical geometry (Vlahakis \& Tsinganos, 1998). 
The general analysis of Paper 1 contains the solution class applied
here, as well as the prominence and loop models by Kippenhahn \&
Schl\"uter~(1957), Hood \& Anzer~(1990), Tsinganos et al.~(1993) and 
Del Zanna \& Hood~(1996). Basically, in this method and under certain 
assumptions, the full MHD equations can be reduced to a system of ordinary 
differential equations (ODE's) which can be integrated by standard methods.

The {\it dynamics} of flows in solar coronal loops may be
described to zeroth order by the well known set of steady
(\(\partial/\partial t=0\)) ideal hydromagnetic equations:
\begin{equation}\label{momentum}
\rho \left( {\bf V}\cdot{\bf\nabla}\right){\bf V}= \frac{1}{4 \pi}
{\left({\bf\nabla}\times{\bf B}\right)\times{\bf B} }
-{\bf\nabla}P-\rho g \hat Z \,,
\end{equation}
\begin{equation}\label{fluxes}
\bf{\nabla}\cdot{\bf B}=0\,,\quad \bf{\nabla}\cdot\left(\rho{\bf
V}\right)=0 \,,\quad \bf{\nabla}\times\left({\bf V}\times{\bf
B}\right)=0 \,,
\end{equation}

\noindent where \({\bf B}\), \({\bf V}\), \(-g \hat Z \)  denote
the magnetic, velocity and (uniform) external gravity fields while
\( \rho\) and $P$ are the gas density and pressure. The {\it
energetics} of the flow on the other hand is governed by the first
law of thermodynamics :
\begin{equation}\label{firstlaw}
q=\rho {\bf V} \cdot \left [ {\bf\nabla} e+P {\bf\nabla} \frac{1}{\rho} \right
]
=\rho {\bf V} \cdot \left [ \bf{\nabla} h-\frac{1}{\rho}\bf{\nabla} P \right ]
\,, \label{qdef}
\end{equation}
where $q$ is the net volumetric rate of some energy input/output,
$\Gamma=c_{p}/c_{v}$ with $c_{p}$ and $c_{v}$ the specific
heats for an ideal gas, and

\begin{equation}
e=\frac{1}{\Gamma-1}
\frac{P}{\rho}
\end{equation}

\noindent the internal energy per unit mass, with $h=\Gamma e$ the
corresponding  enthalpy function.

At present, a fully three-dimensional MHD equilibrium modelling with
compressible flows is not amenable to analytical treatment and so we 
assume translational symmetry. Thus, we assume that in Cartesian 
coordinates $(Z, X, Y)$, the 
coordinate $Y$ is ignorable (\(\partial/\partial Y=0\)) and the magnetic
and flow fields are confined to the $Z$ - $X$ plane. 
We model the profile of the loop in the $x$-$z$ plane and ignore 
variations in the $y$-direction, i.e., we assume that the physics of the 
$x$-$z$ plane is independent of what happens across the loop in the 
$y$-direction.  All previous equilibrium models of coronal loops mentioned 
above have been one-dimensional and non-magnetic.  To begin with, we  
represent $\bf B$ by using a magnetic flux function
(per unit length in the $\hat{Y}$ direction)

\begin{equation}
{\vec B}=\vec \nabla A(Z,X)\times\hat{Y} .
\end{equation}

\noindent Then, there exist free integrals of $A$ including the ratio of the
mass and magnetic fluxes on the poloidal plane (Z-X), $\Psi_A(A)$,

\begin{equation}
{\vec V}={{\Psi}_A\over 4\pi\rho}{\vec B}
\,,
\end{equation}
where the stream function $\Psi $ is a function of the magnetic flux
function $A$ and ${\Psi}_A$ is its derivative (Tsinganos, 1982).  The component
of Eq. (\ref{momentum}) along the field may be written as

\begin{equation}
\rho {\bf V} \cdot\nabla I=0,\label{alignedmom}
\end{equation}

\noindent where

\begin{equation}
I=I(A)=\frac{V^2}{2} +gz + \int_{s_0}^s \frac{1}{\rho}
\frac{\partial P}{\partial s} ds
\end{equation}

\noindent is the generalised classical Bernoulli integral, a further integral
of the flow.  Eqs. (\ref{qdef},\ref{alignedmom}) may be added to describe the
momentum balance

\begin{equation}
q=\rho {\bf V} \cdot\nabla E\,,
\end{equation}

\noindent in terms of $E$, the total energy of the flow

\begin{equation}
E=\frac{V^2}{2} +gz + h.
\end{equation}

\noindent In general, because of the heat source $q$, the total energy is not
conserved along the loop (Sauty \& Tsinganos, 1994).  Even in the polytropic
case where the pressure takes the special form $P=Q(A)\rho^{\gamma}$ the net
volumetric rate of energy in/out

\begin{equation}
q=\frac{\gamma -\Gamma}{\Gamma -1}\frac{P}{\rho}{\bf V}\cdot\nabla\rho
\end{equation}

\noindent is not generally zero (Tsinganos et al.,~1992).  Only in the special
polytropic case with $\Gamma =\gamma$ is the flow adiabatic, and the total
energy coincides with the generalised Bernoulli integral
and is conserved.  However, the general non-polytropic case is the case of
interest in this paper.

In the general case, the system of equations (\ref{momentum}) -
(\ref{fluxes}) should be solved simultaneously with a detailed energy balance
equation 
in order to yield a self-consistent calculation of the equilibrium 
values of $\rho$, $P$, $\bf V$ and $\bf B$ along the loop. However, it is 
a fact that the detailed forms of the several heating/cooling mechanisms in the
energy equation are 
not known, e.g., we do not know the exact expression of the heating along 
coronal loops which contributes, among others, to the various parts of the 
net volumetric heating rate q in Eq. (3). Hence, a compromising strategy is to
use, 
for example, a polytropic equation of state and then solve for the  
values of $\rho$, $P$, $\bf V$ and $\bf B$.       
Then, we may determine the volumetric rate of net heating  
from Eq.(\ref{firstlaw}). In such a treatment the heating sources
which produce some specific solution are not known {\it a priori};
instead, they can be determined only {\it a posteriori}. In this
paper we shall follow a similar approach.
\subsection{The solution class}
\label{thesolutionclass}

In order to proceed to the analytical construction of some classes of exact 
solutions for coronal loops, we make two key assumptions:

 \begin{enumerate}
  \item that the Alfv\'en number $M$ is solely a function of the
dimensionless horizontal distance $x=X/ Z_0$, i.e.,

\begin{equation}\label{Alfven}
M_{}^{2}=\frac{4 \pi\rho V^{2}}{B^{2}}=\frac {\Psi_{A}^{2}}{4 \pi 
\rho} = M^2(x)\,, \label{M2}
\end{equation} 
and
  \item that the velocity and magnetic fields have an exponential 
dependence on $z=Z/Z_0$,
\begin{equation}\label{assumptions2}
A= Z_0 B_{0} {\cal A}\left(\alpha\right)\,,
\qquad
\alpha=G(x) \exp{(-z)}
\,,\label{G}
\end{equation}
\end{enumerate}

\noindent
for some function $G(x)$, where $Z_0$ and $B_0$ are constants.  
With this formulation the magnetic field has the form

\begin{equation}
{\vec B}= B_0 \alpha {\cal A}^{'}(\alpha) \left[\hat{X}+F(x) \hat{Z}\right] 
\,,
\end{equation}
where
\begin{equation}
F(x) = \frac{1}{G(x)}\frac{dG(x)}{dx}  = \left(\frac{dZ}{dX}\right)_A \label{F}
\end{equation}
is the slope of the field line.  This is the analogue in Cartesian geometry of
the "expansion factor" in the related wind models in spherical geometry (see
Sauty \& Tsinganos, 1994).  The function $G(x)$ also has a physical meaning. 
Either by integrating Eq. (\ref{F}) or by inverting Eq. (\ref{G}) the equation
for the field line defined by $\alpha =\alpha_0$ is found to be

\begin{equation}
z=\log G(x)-\log\alpha_0 .\label{fieldlineeq}
\end{equation}

\noindent This is the Cartesian analogue of the cylindrical distance of a
field line from the polar axis in spherical wind theory (see Sauty \&
Tsinganos, 1994).  With these assumptions, the
momentum-balance equation may be broken down into a system of
first-order ODE's for functions of $x$, and a corresponding system
of ODE's for corresponding functions of the magnetic flux function.
The methods of obtaining these ODE's are described in Paper 1, where all 
existing solutions are summarised in Table 1 therein. 

The solutions used in this paper are taken from the first family in Table 1 in
Paper 1.
In the remaining of this section we examine the general case, with all
constants non-zero.  The
corresponding expressions for the magnetic flux function $A$, the mass flux
per unit magnetic flux $\Psi_A$, the density $\rho$, the magnetic induction
$\bf B$ and the velocity $\bf V$ are (see Paper 1)

\begin{eqnarray}
A(\alpha ) & = & Z_0 B_0 \int\sqrt{2C_1 +\lambda C_2 \alpha^{\lambda
-2}}d\alpha ,\\
\Psi_A (\alpha ) & = & \frac{B_0}{\sqrt{gZ_0}} \sqrt{2D_1 \alpha^2 +\lambda
D_2\alpha^{\lambda}} ,\\
\rho (x,\alpha )& = & \frac{B_0^2}{4 \pi g Z_0} \frac{2D_1 \alpha^2 +\lambda
D_2\alpha^{\lambda}}{M^2} ,\\
P(x,\alpha ) & = & \frac{B_0^2}{4\pi} \left[ P_0+P_1(x) \alpha^2+ P_2(x)
\alpha^{\lambda}\right] ,\label{pressuresum}\\
{\vec B} & =& B_0 \sqrt{2C_1 \alpha^2 +\lambda C_2\alpha^\lambda} 
\left[\hat{X}+F(x) \hat{Z}\right] ,\\ 
{\vec V} & =& \sqrt{g Z_0} \sqrt{\frac{2C_1 \alpha^2 +\lambda
C_2\alpha^{\lambda}}{2D_1 \alpha^2 +\lambda D_2\alpha^{\lambda}}} M^2
\left[\hat{X}+F(x) \hat{Z}\right] .
\end{eqnarray}

\noindent where $C_1$, $C_2$, $D_1$,  $D_2$ and $\lambda$ are constants.  Note
that we may choose the constants such that $B_0$ is the  component of the
magnetic field at a 
reference point. Also note that the general non-polytropic case has two 
``scales'': $Z_0$ and $2 Z_0/\lambda$. 
In the expression for the pressure $P_0 = f_0 =$ constant, while 
$P_1$ and $P_2$ satisfy the following ODE's

\begin{eqnarray}
P_1&=&C_1 \left[FM^{2'}- F^{'} (1-M^2)-F^2-1\right]+\frac{D_1}{M^2} \,,
\nonumber \\
P_2&=&C_2 \left[FM^{2'}- F^{'}
(1-M^2)-\frac{\lambda}{2}(F^2+1)\right]+\frac{D_2}{M^2} \,.\nonumber
\end{eqnarray}

\noindent Using the above definitions for the pressure ``components'' together
with the ODE's from Table 1 in Paper 1, we calculate that for
the general case we have the following final system of equations for the
unknown functions of $x$, including the slope of the field lines $F$ :

\begin{eqnarray}
\frac{d \ln G}{dx} & = & F\,, \label{Gd}\\
M^{2'} (x)& = & \frac{\mathcal{C}\lambda F/M^2 - 2F(F^2
+1+P_1/C_1)}{\mathcal{C}/M^4
+ 2}\,,\ \mbox{where}\ 
\ \mathcal{C} = \frac{D_2/C_2-D_1/C_1}{1-\lambda /2}\,, \label{M2d}\\
F ' (x)& = & \frac{FM^{2'} - F^2 -1 + D_1/C_1 M^2 -P_1/C_1}{1-M^2} \,,
\label{Fd}\\
P_1 ' & = & -\frac{2 D_1 F}{M^2} - 2 C_1 (1+F^2 ) M^{2'} - 2 M^2 FF'
\,.\label{P1d}\\
P_2 (x) & = & C_2 \left( F M^{2'} -F'(1-M^2 ) - \frac{\lambda}{2} ( 1+F^2 )
\right)
+\frac{D_2}{M^2 }\,, \label{P2}
\end{eqnarray}

Finally, consider the energy balance along the loop; the net volumetric rate
of heating input/output $q$, equals to the sum of the net radiation $L_R$, the
heat
conduction energy $\nabla\cdot{\bf F}_C$, where ${\bf F}_C$ is the
heat flux due to conduction, and the (unknown) remaining heating
$E_H$,

\begin{equation}
q=E_H+L_R-\nabla\cdot{\bf F}_C.
\end{equation}

\noindent The net heat in/out $q$ is calculated from the MHD model using the
first law of thermodynamics Eq. (\ref{firstlaw}), while the
radiative losses from the optically thin plasma $L_R$ are described by the
equation

\begin{equation}
L_R=-(n/2)^2Q(T),
\end{equation}

\noindent (Raymond \& Smith, 1977) with standard solar atmospheric abundances
as in Rosner et al.~(1978), where $n$ is the particle number density
(we assume
that the plasma is fully ionised) and $Q(T)$ is a piecewise function of $T$
described in
Rosner et al.~(1978).  The thermal conduction
energy is calculated assuming that conduction is mainly along the field, using
the expression

\begin{equation}
-\nabla\cdot{\bf F}_C =  \frac{\partial}{\partial s}\left(\kappa_{\small
||}\frac{\partial T}{\partial s}\right) -\frac{\kappa_{\small
||}}{B}\frac{\partial B}{\partial s}\frac{\partial T}{\partial s}
,\label{conduction}
\end{equation}

\noindent (Spitzer, 1962) where subscripts $||$ indicate values and derivatives
along the
field line, and the variation of the magnetic field strength along the field
line is taken into
account (Priest, 1982, p86).

We present the physical parameters of each loop as functions of arc length
$s$. 
The arc-length along a loop is given by

\begin{equation}
ds^2=dx^2+dz^2=(1+F(x)^2) dx^2
\label{arclength}
\end{equation}

\noindent and the $(x,\alpha )$-dependent physical parameters of a loop can be
understood as functions along the loop by holding $\alpha$ constant (the
definition of a field line 
since $\alpha$ is a flux function) and integrating Eq.~(\ref{arclength}) for
$s$ from the left foot point to the right foot point.

\subsection{Construction of solutions}
\label{construction}

We generate loop-like solutions as follows.  We begin by calculating the right
half of the loop, beginning from the loop apex at $x=0$.  The symmetry
properties of Eqs. (\ref{Gd} - \ref{P2}) ensure that
on integrating from $x=0$ in the negative direction the other half of a
symmetric loop-like solution is obtained.
In the sub-Alfv\'enic case the equations have no critical points and can be
integrated without difficulty.  In the trans-Alfv\'enic case a shooting
algorithm is required to integrate through the critical Alfv\'en point
(Vlahakis \& Tsinganos, 1998; Paper 1) but since steady
super-Alfv\'enic flows have not been observed in the solar atmosphere we will
concentrate on sub-Alfv\'enic examples here.  In this paper we use a similar
shooting algorithm to fix the foot point separation of each sub-Alfv\'enic
loop.  The solution class allows us to fix all physical quantities  at the
apex: the height, magnetic field strength, velocity, density and temperature. 
Having chosen values for these quantities at the apex we begin the
integration.  As the solution approaches the solar surface at $z=0$ it will be
clear whether the foot point separation is greater or less than the desired
(observed) value and a remaining free parameter can be adjusted accordingly. 
This
process is repeated until the solution is fitted to the desired (observed)
configuration.

In this paper we present models fitted to data, where available, in five ways:
we fit the loop height and foot point separation as described above, the
plasma density and temperature, the line-of-sight velocity or velocities of
proper motions whose components perpendicular to the line of sight can be
measured, and we forward-fit synthetic emission models to observed emission
patterns.

It can be seen from the equation for a magnetic field line, Eq.
(\ref{fieldlineeq}), that two field lines defined by $\alpha =\alpha_1$
and $\alpha =\alpha_2$ differ from each other only by a vertical
translation, and that for any point $(Z_1 ,X)$ on the first field line,
the corresponding point on the second field line $(Z_2 ,X)$ can be found
from it by moving vertically a distance

\begin{equation}
Z_2 -Z_1 =\log{\frac{\alpha_1}{\alpha_2}} .
\end{equation}

We may model the cross-sectional width of a loop by taking two such field
lines and by considering the area between these lines to constitute the
loop model.  Then the loop necessarily has maximum cross-sectional width 
at the apex, the remainder of the width profile being uniquely defined by
the geometry of a field line.  Thus the loop width $W$ is not a free
function to be imposed as in one-dimensional studies, e.g., Cargill \& Priest
(1982); Aschwanden \& Schrijver (2002), but is related
to the slope of the loop $F$ by

\begin{equation}
W(s)=\left. \log{\frac{\alpha_1}{\alpha_2}} \right/ \sqrt{1+F^2} .
\end{equation}

If a loop is observed to be tilted with respect to the vertical direction
then we may still model the loop in the $x$-$z$ plane by tilting our
coordinate system accordingly. We must take into
account the effect of this tilt on the physics of the loop.  The
gravitational force acts at an angle to the $z$-axis and the loop cuts
through the stratified atmosphere at an angle.  Therefore in the model the
gravitational force must be multiplied by the cosine of the angle of tilt
and vertical scale heights must be divided by this cosine.

It is well-known that plasma flow is generally present in loops (e.g. Dara et
al.,~2002).  However, only limited information about the magnitude of the loop 
plasma velocities is available today from satellite data: line-of-sight 
measurements from dopplergrams in the case of the
CDS and SUMER data sets, and high-resolution movie measurements of velocities
of inhomogeneities, or proper motions, in the plasma flow in the case of the
TRACE example.  We model
these measurements by taking the two-dimensional velocity field from our
MHD solution and, taking the geometry of the loop and the viewing angles
of the instrument relative to the loop into account, we calculate model
line-of-sight and perpendicular velocity components
to be compared to the observations.  Thus, taking the
planar loop to be confined to the $x$-$z$ plane and centred at the
origin, we define
by $\theta$ the angle in the $x$-$y$ plane between the $x$ axis and the
line from the origin to the instrument, and by $\phi$ the angle between the
line from the origin to the instrument and the plane $z=0$.   Then, assuming
that the distance from the instrument to the loop is much larger
than the size of the loop, the line-of-sight velocity $V_{LOS}$ as observed by
the instrument and the velocities in the two directions of the image
perpendicular to the line of sight, $V_{\perp x}$ and $V_{\perp y}$, are given
by

\[ \left( \begin{array}{c}
V_{\perp x} \\ V_{\perp y} \\ V_{LOS} 
\end{array} \right)
= \left( \begin{array}{ccc}
\cos\theta\cos\phi & \sin\theta\cos\phi & -\sin\phi \\
-\sin\theta & \cos\theta & 0 \\
\cos\theta\sin\phi & \sin\theta\sin\phi & -\cos\phi
\end{array} \right)
\left( \begin{array}{c}
V_x \\ V_y=0 \\ V_z
\end{array} \right) . \] 

\noindent The velocity perpendicular to the line-of-sight has magnitude
$V_{\perp} =\sqrt{V_{\perp x}^2+V_{\perp y}^2}$, so that
$V^2=V_{\perp}^2+V_{LOS}^2$.

\section{Observations, data reduction and loop diagnostics}
\label{observations}

There is some controversy surrounding the issue of extracting measurements of
coronal densities and temperatures from emission data.  Judge \&
McIntosh~(1999)
contrast the probable multi-thermal nature of loops
consisting of strands with inefficient cross-field thermal conduction (Litwin
\& Rosner,~1993) with the evidence from TRACE that loops in
significantly different temperature filters are never co-spatial, and stress
the
ill-posedness and non-uniqueness of inverse modelling techniques applied to
the transition region and corona.  In this work, densities and
temperatures have been calculated for the TRACE example using the
the narrowband 171~\AA\  and 195~\AA\  two-filter fluxes (e.g., Aschwanden
et al.~2000; Winebarger et al.~2002). Forward-fitting of our model
to two-filter fluxes, $F_{171}(s)$ and $F_{195}(s)$, does not suffer
from the ambiguity of filter-ratio temperature fits,
$R=F_{195}/F_{171}$,
which has been shown to have, besides the $T \approx 1.0$ MK solution,
also a
high-temperature solution at $T\approx 5.0$ MK (Testa et al.~2002).
But Winebarger et al.~(2002) demonstrated that the $T=5.0$ MK
solution of Testa et al.~(2002) is generally not consistent with
combined
TRACE and Yohkoh/SXT data, and similarly, Chae et al.~(2002)
demonstrated
that the $T=5.0$~MK solution is not consistent with TRACE triple-filter
data.
   An additional confusion in the temperature analysis of multi-filter
data
was raised by Schmelz et al.~(2001), who showed that the emission
measure
distribution of a loop structure observed with CDS over a temperature
range
of $\log T=5.4-6.4$ displays a rather broad temperature distribution
with
the mean temperature increasing towards the loop top, and thus concluded
that the analysed CDS loop structure has at every location a broad
temperature
distribution and heating occurs at the loop top. Martens et al. (2002)
characterised the smoothed DEM of Schmelz et al.~(2001) as
a flat plateau and pointed out that any filter-ratio method is inaequate
to
determine the temperature of such a loop system (see also Schmelz,~2002).
However, the CDS
observations
of Schmelz et al.~(2001) can easily be understood if the following facts
are taken
into account: (1) The effective spatial resolution of CDS is $\approx
10"-15"$,
compared with $\approx 1"$ of TRACE, (2) TRACE 171~\AA\  images reveal
for every loop
structure observed with CDS at $T\approx 1.0$ MK at least $\approx 10$
loop
threads, (3) the broad DEM distribution of a CDS loop structure is not
smooth
but consists of multiple temperature peaks which clearly indicate
multiple loop
threads with different temperatures (Aschwanden~2002), (4) the centroid
position
of the CDS loop structure was found to exhibit displacements in each CDS
temperature filter (as presented by Trae Winter at the "Coronal Loop
Workshop"
in Paris, November 2002), which confirms that the CDS loop structure
consists
of multiple, non-cospatial loop threads, and (5) the combined emission
measure distribution of many loop threads over a broad a temperature
range
bears a hydrostatic temperature bias that yields an average temperature
increasing with height (Aschwanden \& Nitta~2000). From these facts
there is
clear evidence that a loop structure seen by CDS consists of multiple
loop
threads with different spatial positions and different temperatures although
Martens et al.~(2002) argue that, because
the high-temperature edge as well as the low-temperature edge of their DEM
plateau moves towards higher temperatures approaching the loop top, high
temperatures must exist near the loop top which are not found lower down the
structure, and so individual loop strands are unlikely to be exactly
isothermal.  Of
these loop threads, TRACE
resolves a subset that coincides in the temperature sensitivity range of
a TRACE
narrow-band filter. It is therefore imperative to apply a
model only to a resolved loop thread, rather than to a multi-temperature
bundle
of loop threads that make up a CDS loop structure. Since we analyse the
same
loop structure as described in Schmelz et al.~(2001), we apply our MHD
model only to a single CDS temperature filter (Mg~{\sc ix} 368~\AA ,1~MK),
being aware that
even the
loop structure seen in this single filter still consists of multiple
threads,
given the poor CDS resolution, and thus expect only to extract average
density and velocity parameters for this loop system at
the given temperature range of the Mg~{\sc ix} filter ($T=1$~MK). Also, we
apply a
forward-fitting technique to the observed emission, as recommended by
Judge \& McIntosh~(1999), to avoid the ill-posedness of filter-ratio
techniques.
In the three cases we model here, we use in our forward-modeling only a
single image (CDS, SUMER) or an image pair of similar temperature
(TRACE) to avoid confusion between loop strands of different
temperatures.

\begin{figure*}
\begin{center}
\resizebox{0.49\hsize}{!}{\includegraphics*{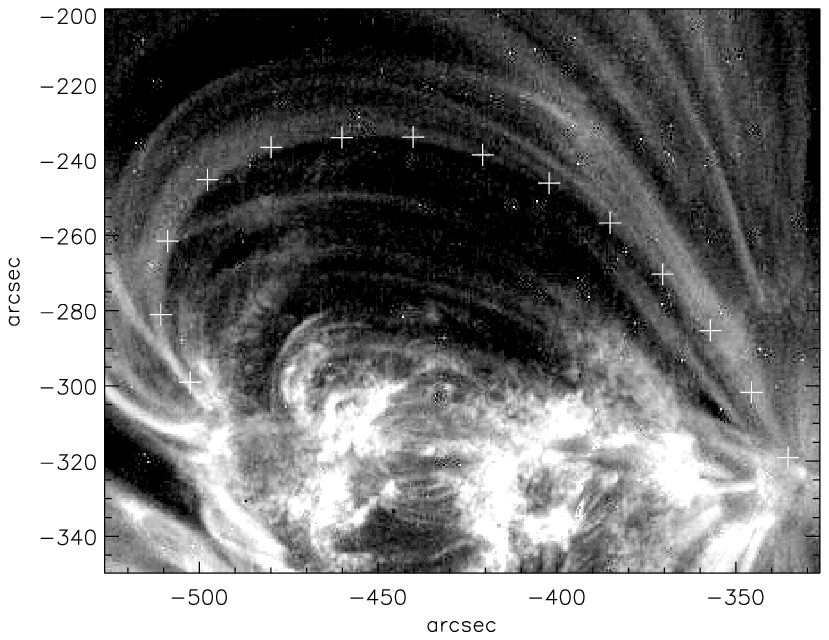}}
\resizebox{0.49\hsize}{!}{\includegraphics*{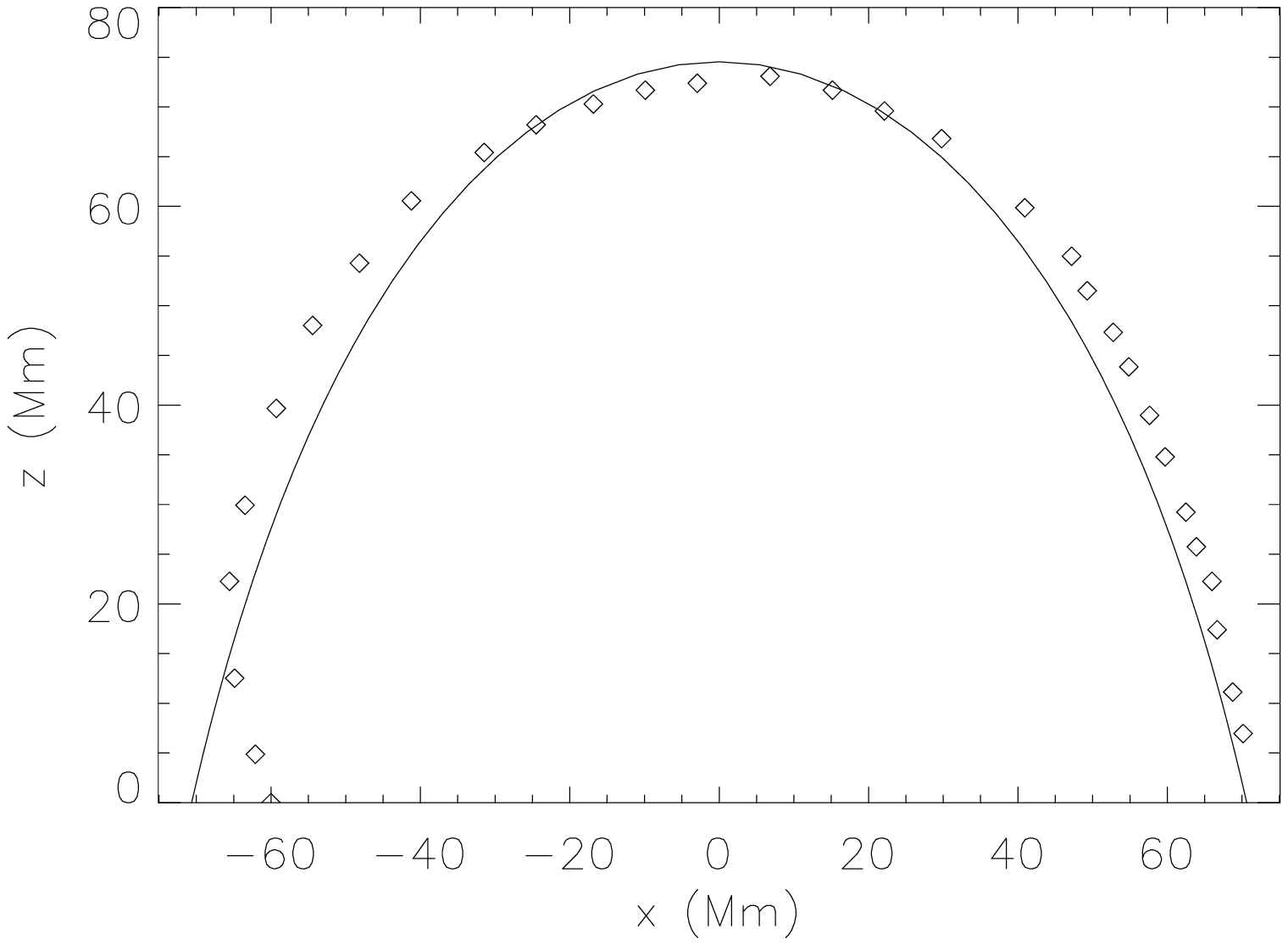}}
\end{center}
\caption{An MHD model of a loop observed by TRACE fitted to
observational data: shown are the TRACE image of the loop system
with the loop of interest contained within crosses (left
picture) and the model field line (solid line) fitted to the observed
line, represented by diamonds (right picture) in the $x$-$z$ plane.}
\label{loop2pics}
\end{figure*}

We use observations from TRACE in the 195~\AA  and 171~\AA\   
bands taken on 24-26 October 1999, SoHO CDS observations used
by Schmelz et al.~(2001) taken on 20 April 1998 and SUMER observations from
March 25, 1996.  
The TRACE instrument was pointing on a
well-defined isolated loop system at $-426\arcsec,-275\arcsec$
(see Fig.~\ref{loop2pics}, top left picture). The field of view is
of $768 \times 768 $ pixels whereas the pixel size is 0.5\arcsec.
The corrections that we applied are the following: we subtracted
the readout pedestal and the dark current, we cleaned out the
pixels damaged due to cosmic-rays and we extracted the CCD readout
noise.

In order to derive the geometry of the loop as well as the
physical parameters we followed Aschwanden et al.~(1999).  We used the {\sc
stereo}
package (Aschwanden et al.,~1999), which is part of the solar software (SSW) in
order
to reproduce the geometry of the loops. As the lines used are
optically thin, when we diagnosed physical parameters such as the
electron density, we took great precautions to extract the
background emission as in Aschwanden et al.~(1999). We supposed that the
background
around a loop can be derived from a stripe of emission which
contains the loop (see Fig.~\ref{loop2pics}, top left picture).
Each cross-section of a stripe contains a cross-section intensity
profile of the loop as well as non-loop intensity that surrounds
it.
The intensity \lq outside\rq\, the loop, was interpolated across
the loop region in order to simulate the background contribution
there. We tried several boundaries of the loop profile and choose
those maximising the difference between the total and the
background flux, as in Aschwanden et al.~(1999). The extraction of the
background was
performed for the loop as it appears in the Fe~{\sc ix} 171~\AA\
and the Fe~{\sc xii} 195~\AA\ lines. We computed the temperature
and the emission measure using the {\sc trace\_teem} routine which
applies a filter ratio technique with the Fe~{\sc ix} 171~\AA\ and
the Fe~{\sc xii} 195~\AA\ lines. We derived the mean electron
density $n_e$ at each point along the loops using
Eqs.~(\ref{electrondensity}),

\begin{equation}
n_e\,=\,\sqrt{\frac{EM}{w}}\ , \ \ \ w=\frac{ \int F_T( s,t_j) -
F_B(s,t_j)\, dt_j}{ max (F_T( s,t_j) - F_B(s,t_j) )}
\label{electrondensity}
\end{equation}

where $w$ is the loop width, $F_T(s,t_j)$ is the total flux at a
distance $s$ along the loop and at $t_j$ across it, and
$F_B(s,t_j)$ is the background flux.

\begin{figure*}
\begin{center}
\resizebox{0.99\hsize}{!}{\rotatebox{90}{\includegraphics*{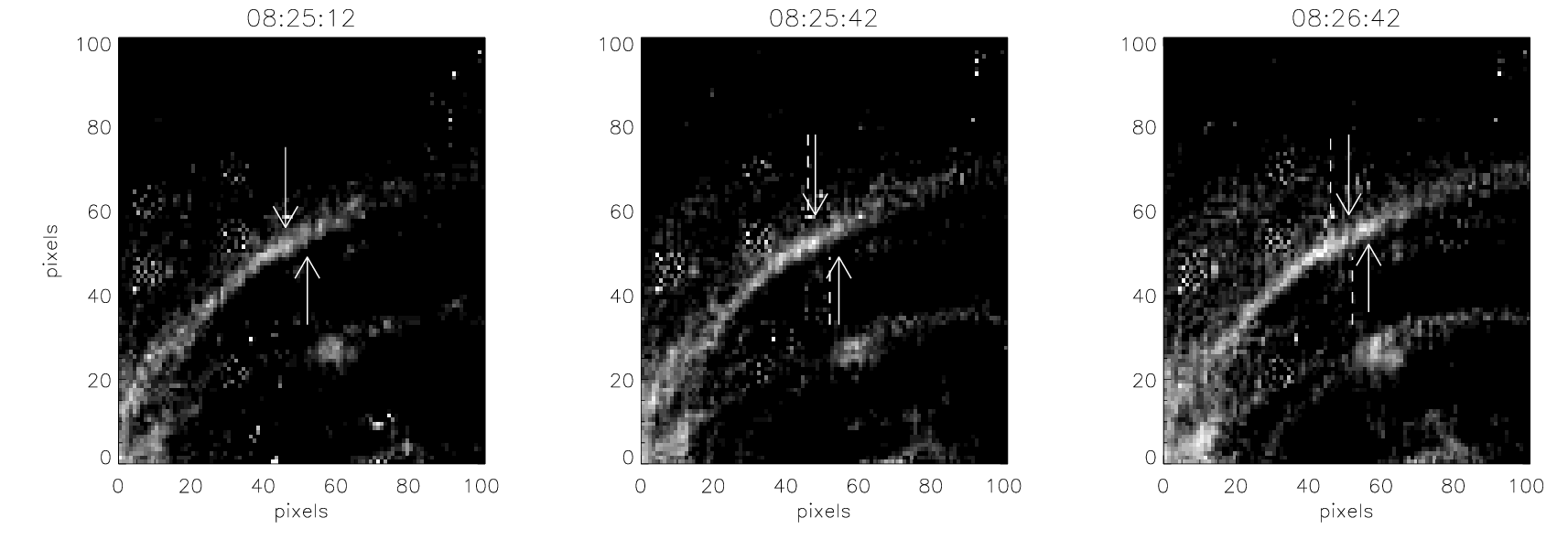}}}
\resizebox{0.49\hsize}{!}{\includegraphics*{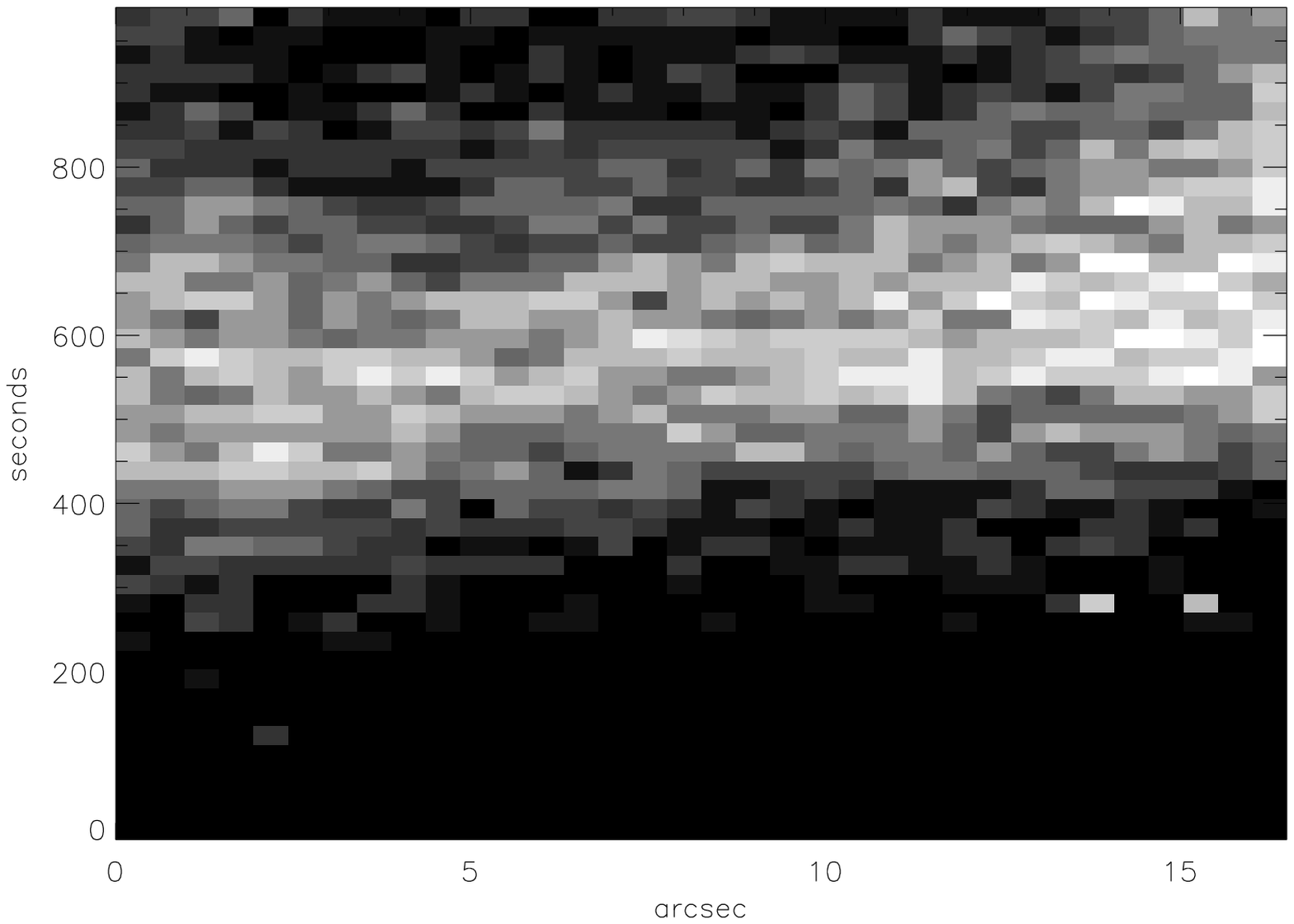}}
\end{center}
\caption{Proper motions along the loop shown in Fig. \ref{loop2pics}. We show
in the top three pictures a
part of the loop 
in three 171~\AA\ images close in sequence, showing the displacement of two
blobs of
material indicated by arrows.
The dashed lines in the second and third pannel show the initial positions of
the 
two blobs. The mean velocity we calculate is of the order of 30 km/sec.  We
show in the bottom picture the evolution of the intensity along another
segment of the loop (horizontal axis) versus time (vertical axis).  Black
represents unenhanced loop emission and shades of grey represent enhanced
emission.
This variation of intensity travelling toward the right footpoint of the
loop may be associated with
a flow along the loop.  We can estimate roughly from this figure a velocity of
40-50 km/sec.}
\label{movie}
\end{figure*}

We tried to measure the proper motions, if any, of the loop plasma.
We first centered very carefully the 171 \AA\ every 30 sec. images and made
a movie with them. Fig. \ref{movie} shows frames from this movie of proper
motions along the loop. We show a part of the loop in three 171~\AA\  images
close in sequence, showing the displacement of two blobs of material indicated
by arrows.  The dashed lines in the second and third pannel show the initial
positions of the two blobs. We believe that the material is moving - or the
excitation is moving - from the left to the right foot point.  Since
half a pixel is the minimum displacement and it corresponds to a velocity of
17 km/s, we consider this as the error of the measurements.  As ``points'' we
select bright features within the loop, which can
be followed in at least two images.  The various
points measured were located in only two images, with the exception of
three points which are each found in three images.  The measurement is very
subjective, but
since quite some points are measured, especially near the top, and most of
their velocities are within the range of 30-40 km/s we believe that this
value is close to the real velocity.  
The mean velocity that we calculate is of the order of  30 km/sec.  We show in
the bottom picture of Fig.~\ref{movie} the evolution of the intensity along
another segment of the loop plotted against time.
This variation of intensity travelling toward the right footpoint of the
loop may be associated with
a flow along the loop.  We can estimate roughly from this figure a velocity of
40-50 km/sec.  It is
possible that the observed ``proper motions'' could also be wave
disturbances, propagating with very slow subsonic speed. To distinguish
between a flow motion and a wave one test would
be:
a flow usually has a temperature difference to the pre-existing plasma
in
the loop, while a wave motion does not change the temperature at all.  For
these observations we lack 195~\AA\  images that are close in time to the
series of 171~\AA\  images where we see the motion and so we cannot check the
temperature variation at the position of the blobs.  However, in the 171~\AA\ 
images we observe individual blobs that we can follow in up to three images.
We do not detect
any oscillation of the intensity in this particular loop.  This as well as the
slow velocities found leads us to conclude that these velocities are
associated with plasma flow and not wave motion.

As for the CDS data, we applied the usual CDS procedures to treat the
geometrical
corrections and to calibrate them. The dopplershifts along the loop, are
computed in
the Mg~{\sc ix} 368~\AA  (1~MK) line. 
For each selected point along the loop, we took the sum of the 4 nearest
individual spectral profiles (corresponding to 
4 spatial pixels). Thus, for each selected point we applied to that less noisy
spectral profile a double gaussian fit to take into account 
the blend due to the
Mg~{\sc vii}
line at 367~\AA.  Before the fitting, we subtracted from each
spectral profile a background one, selected from dark
regions near the loop. In order to estimate what should be the zero velocity,
at the surface of
Sun, we selected an area on the Mg~{sc ix} dopplergam, on the disk, but very
close to the limb.
The wavelength calibration was based on the assumption that the average
dopplershift near the limb
is very close to zero, as it was suggested from works like the one by Peter \&
Judge~(1999).

The SUMER data we used were obtained during a raster that took place on March
25, 1996.
The instrument recorded the Ne~{\sc viii} 770, 780~\AA\, and the  C~{\sc iv} 
1548~\AA\
spectral lines. We applied the usual SUMER data reduction and geometrical
corrections.
We calculated  the dopplershifts along the loop following the same method as
with CDS.
The background spectral profile was also subtracted before the fitting
procedure. As we couldn't use a reference spectrum to calibrate the measured
dopplershifts
(as it is done in e.g. Teriaca et al.,~1999) we selected a quiet Sun area away
from
the active region and we supposed that there should be a blueshifted of
2~km/s, which is the mean measured dopplershift for that line (Peter,~1999;
Dammash et al.,~1999).

\section{The models}
\label{models}

We describe in this section details of the three loops as observed and
modelled.  The models are fitted to the observations in many different ways:
by geometry (loop height, foot point separation and, less precisely, loop
width), emission measure, density, temperature and velocity.  The resulting
momentum balance, energy profile and heating profile are then described.

\subsection{The observable quantities: loop geometry, emission measure,
density, temperature and velocity}

\begin{figure*}
\begin{center}
\resizebox{0.99\hsize}{!}{\includegraphics*{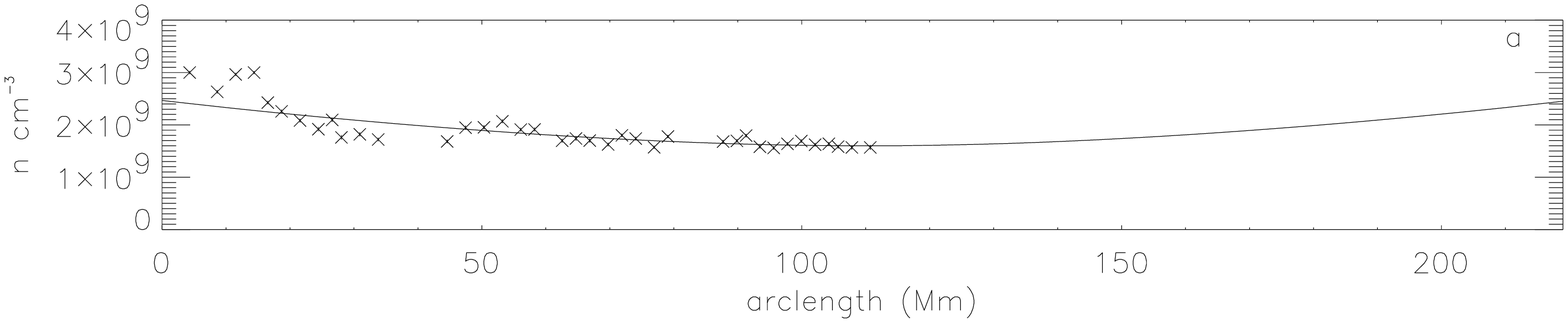}}
\resizebox{0.99\hsize}{!}{\includegraphics*{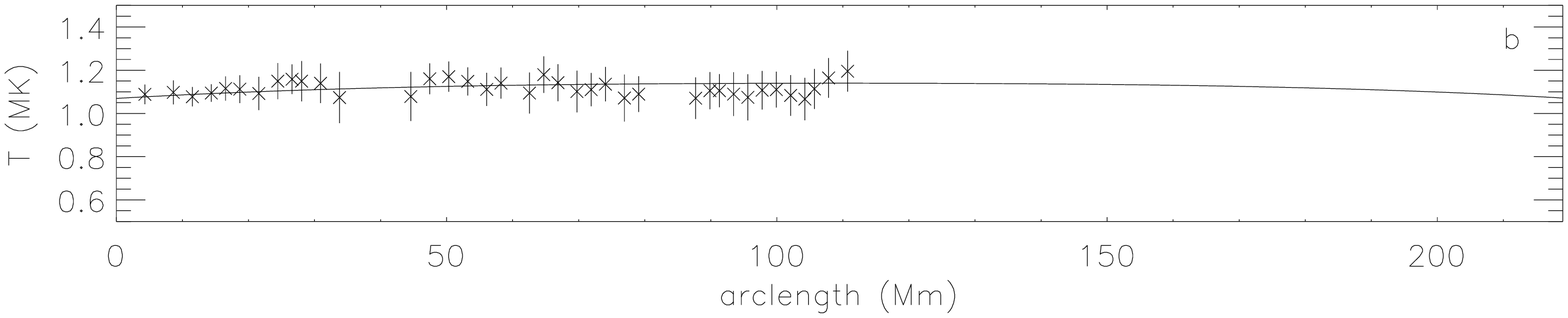}}
\resizebox{0.99\hsize}{!}{\includegraphics*{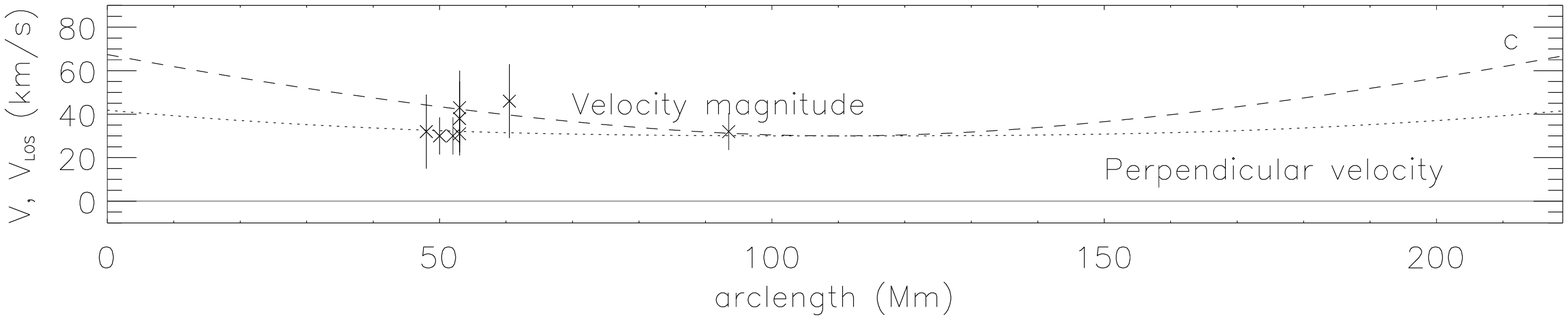}}
\resizebox{0.99\hsize}{!}{\includegraphics*{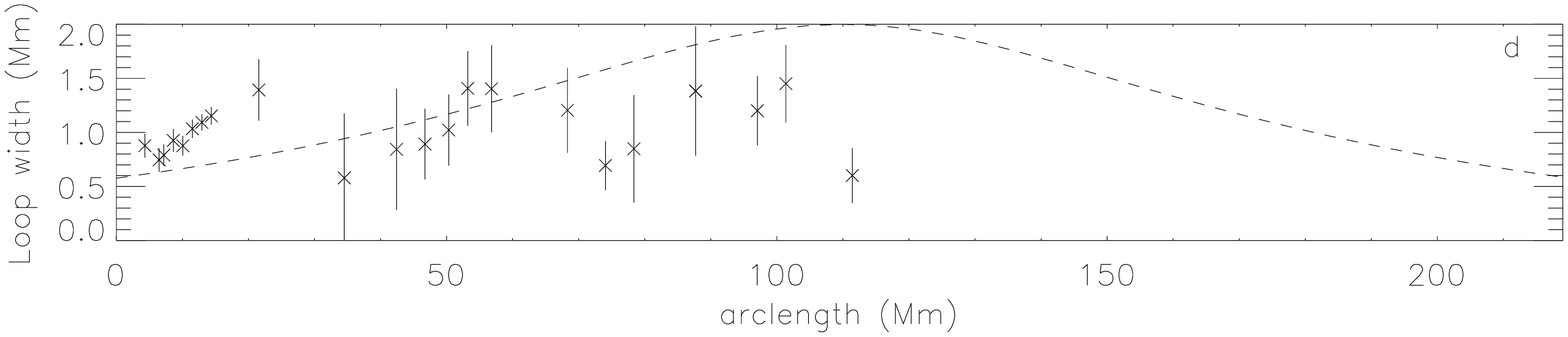}}
\resizebox{0.99\hsize}{!}{\includegraphics*{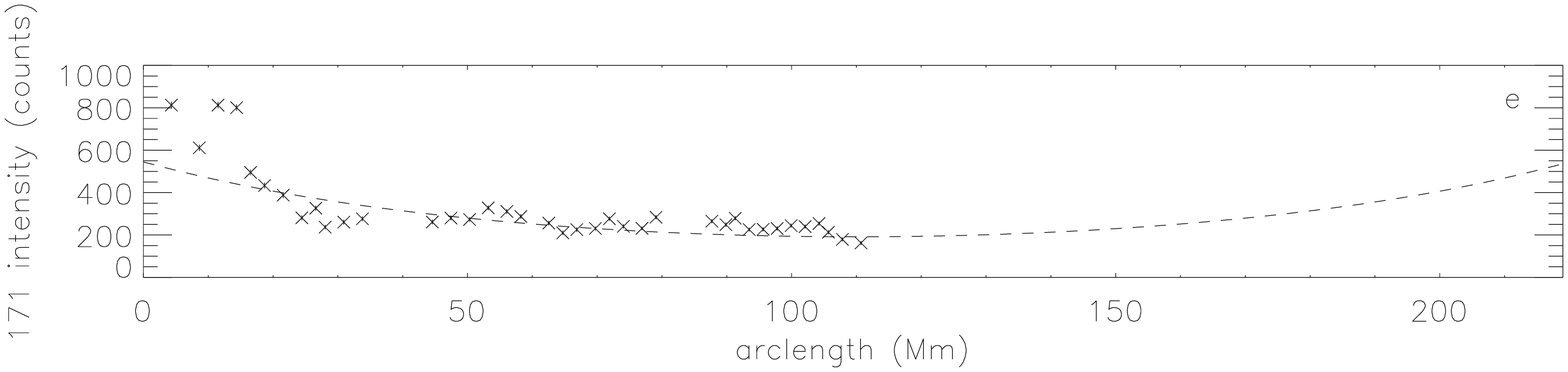}}
\resizebox{0.99\hsize}{!}{\includegraphics*{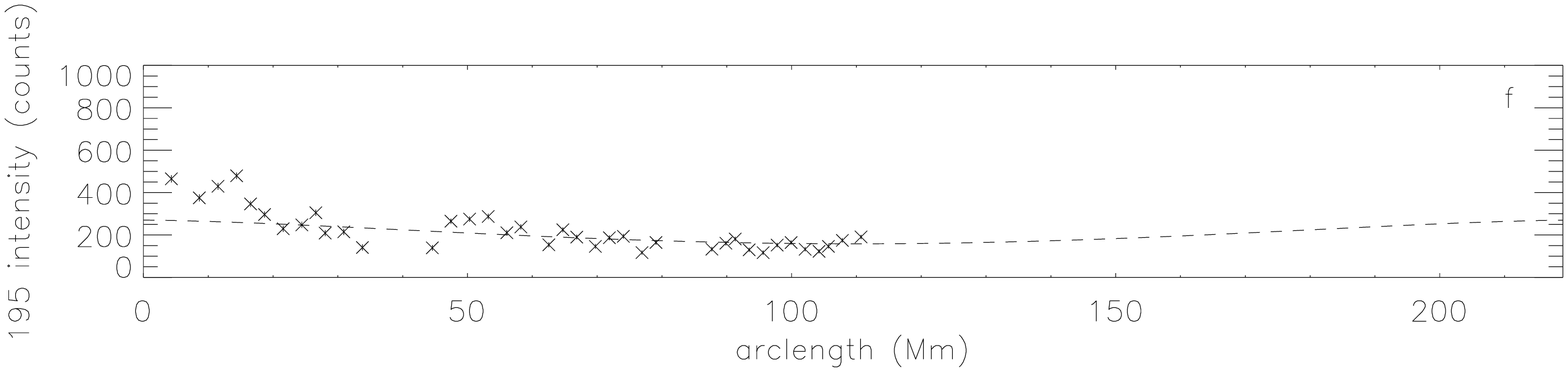}}
\end{center}
\caption{An MHD model of a loop observed by TRACE fitted to
observational data: shown are (a) the particle number density, (b) the
temperature, (c) the flow velocity, (d) the loop width and (e) the 171~\AA\ and
(f) the 195~\AA\ emission patterns compared with the synthetic
forward-modelled emission from the MHD model.  The forward-modelled emission
patterns are computed using the TRACE response functions.  All are graphed
against arc length
along the fieldline of the loop shown in Fig.~\ref{loop2pics}. In
these plots, the observed values are
represented by $x$ symbols and the model by the lines. In the
velocity plot the modulus $|{\bf V}|$ is graphed with a dashed
line while a simulated perpendicular velocity profile is
represented by a dotted line.}
\label{loop2params}
\end{figure*}

\begin{figure*}
\begin{center}
\resizebox{0.49\hsize}{!}{\includegraphics*{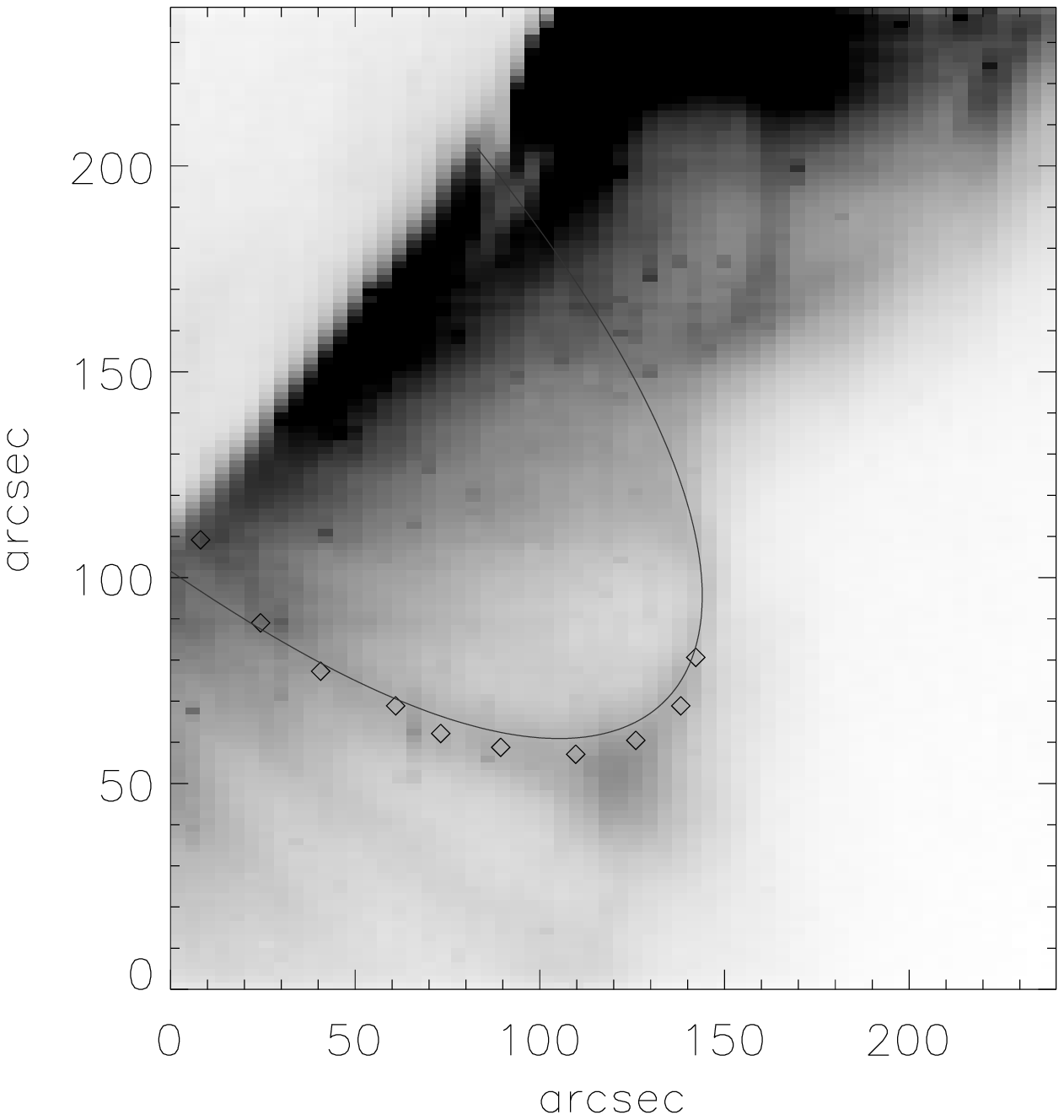}}
\resizebox{0.49\hsize}{!}{\includegraphics*{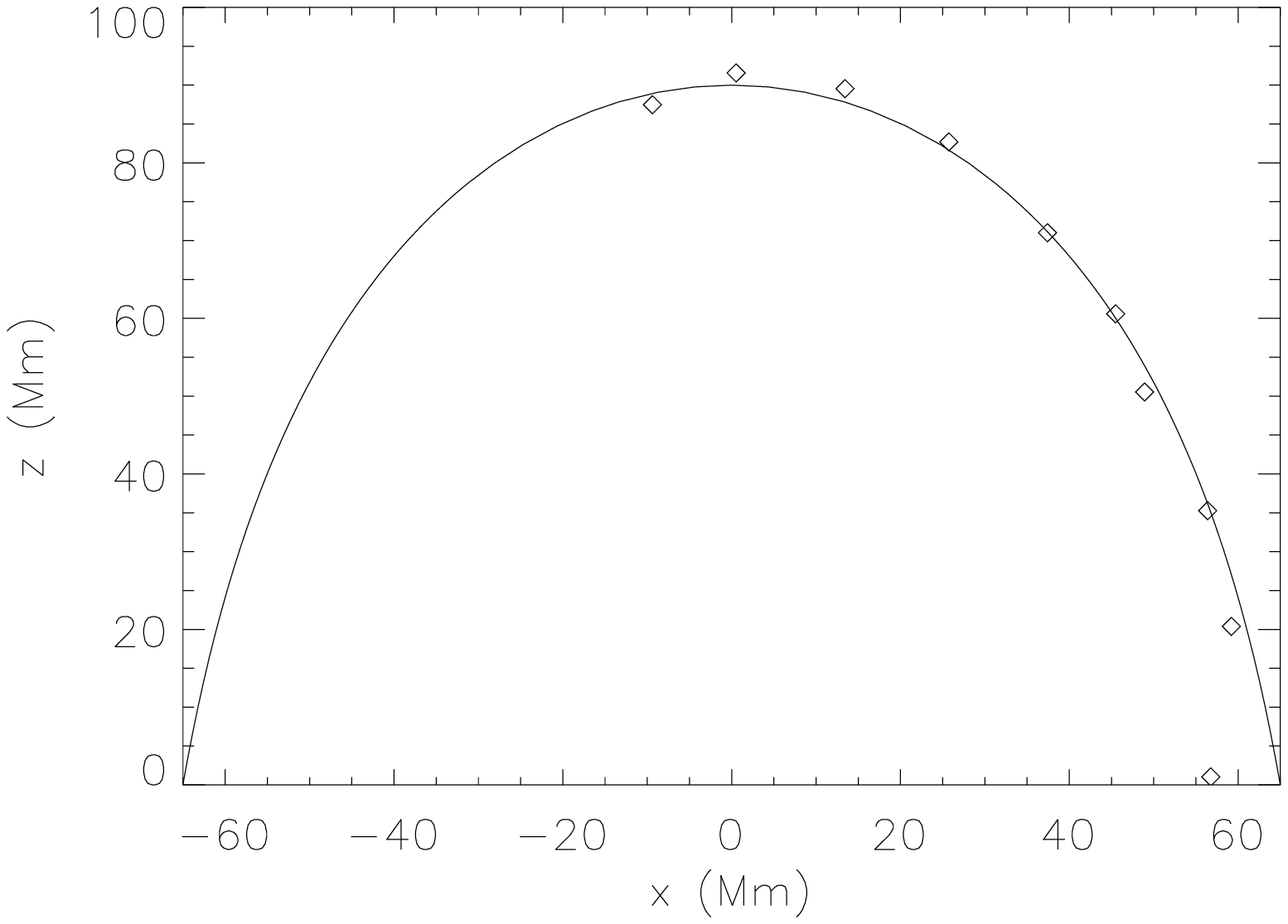}}
\end{center}
\caption{An MHD model of a loop observed with SoHO/CDS (Schmelz et al.,~2001),
fitted to
observational data: the plots are organised as
in Fig.~\ref{loop2pics}.  The CDS image of the loop is in Si~{\sc
xii} 520.66~\AA\ (left picture), and shows the points used for
the loop fitting, as well as the fitted loop
of the right picture projected onto the image. The fitted
field line of the model is shown in the $x$-$z$ plane along with
these points (diamonds) for further comparison.} \label{schmelzpics}
\end{figure*}

\begin{figure*}
\begin{center}
\resizebox{0.99\hsize}{!}{\includegraphics*{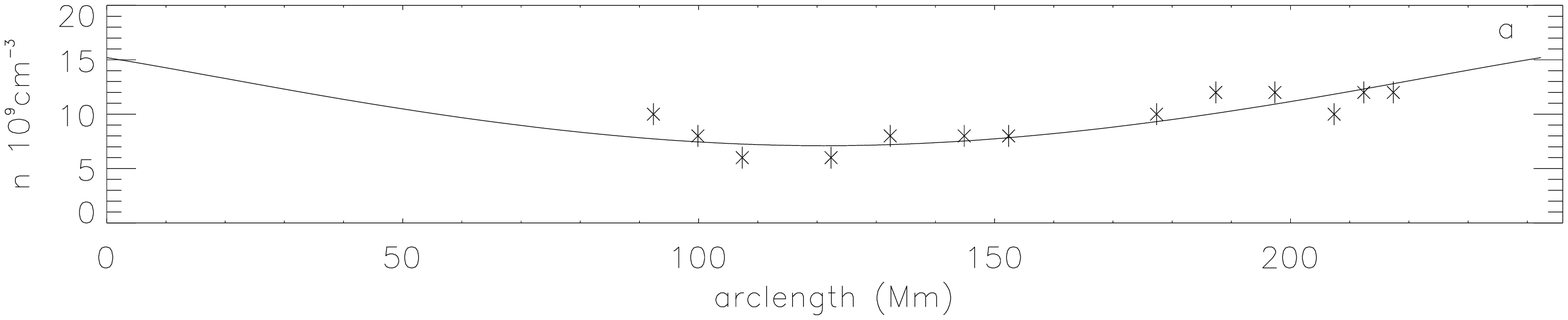}}
\resizebox{0.99\hsize}{!}{\includegraphics*{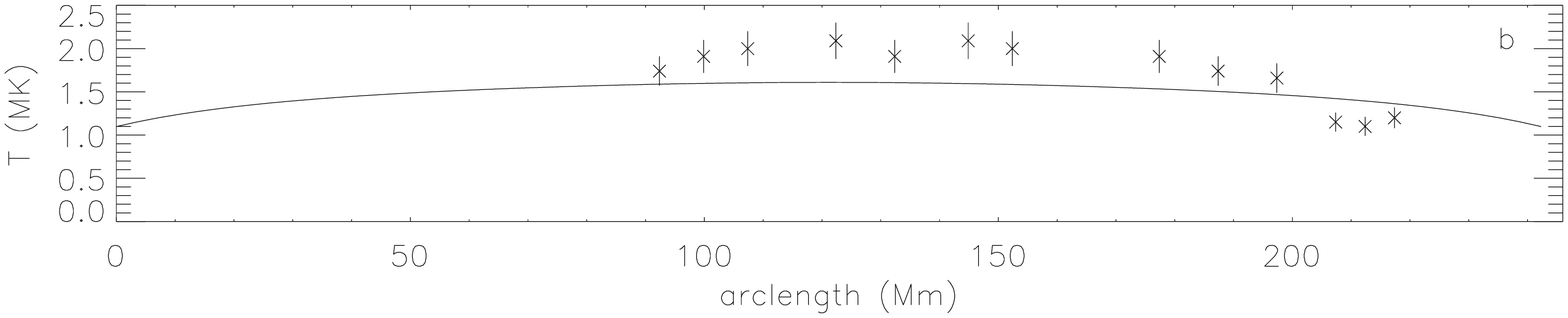}}
\resizebox{0.99\hsize}{!}{\includegraphics*{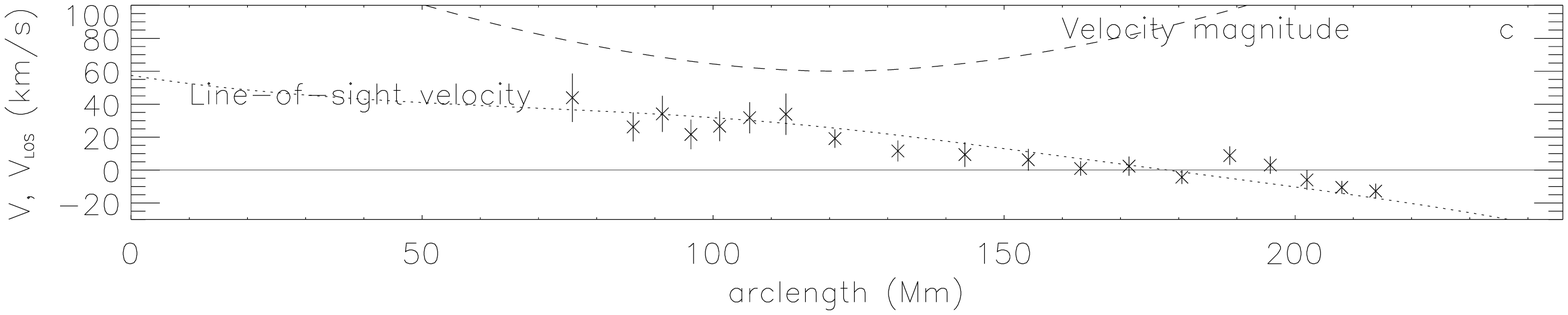}}
\resizebox{0.99\hsize}{!}{\includegraphics*{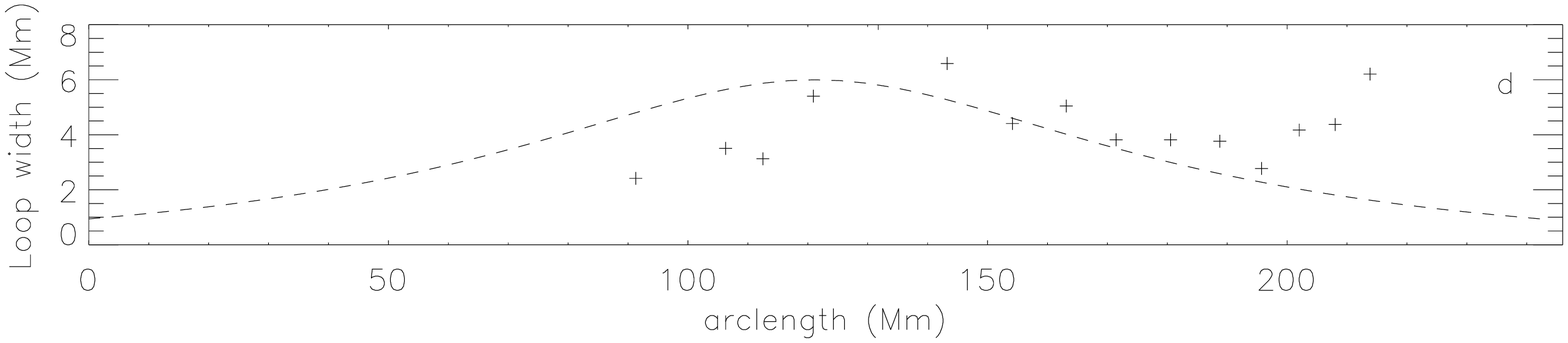}}
\resizebox{0.99\hsize}{!}{\includegraphics*{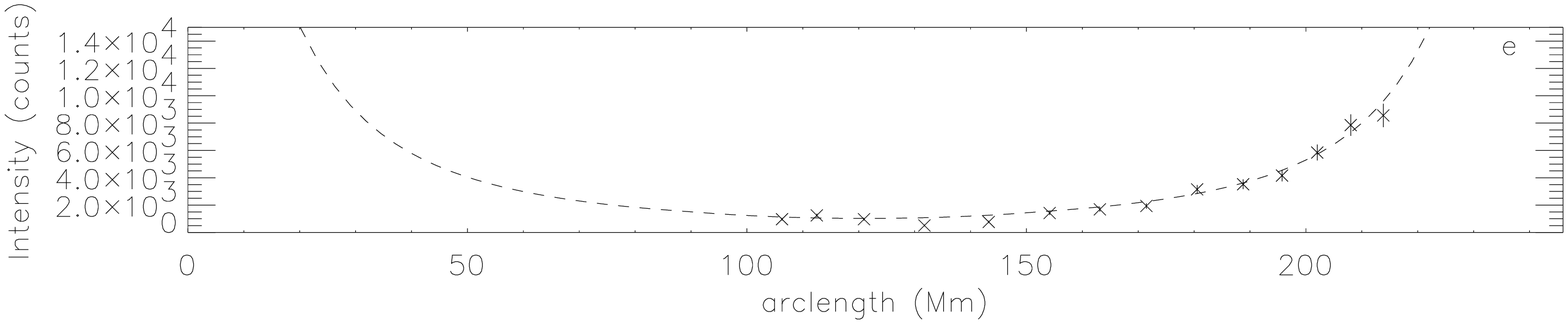}}
\end{center}
\caption{An MHD model of a loop observed with SoHO/CDS (Schmelz et al.,~2001),
fitted to
observational data: the plots are organised as
in Fig.~\ref{loop2params} . The observational
data in the density and temperature plots are taken from Table 2
of Schmelz et al.~(2001).  Because the Si~{\sc xii} emission data from the
image in Fig.~\ref{schmelzpics} and the multi-thermal DEM density and
temperature data are inconsistent, priority has been given to fitting the
emission model, while the density and temperature data are used to guide the
modelling (see text).  The result is a near-isothermal model.  The
observational data in the velocity plot were calculated from dopplergrams
computed from the same Si~{\sc xii} image in Fig.~\ref{schmelzpics} using
Gaussian fitting techniques. Error bars are omitted from the width plot because
the errors are too large.} \label{schmelzparams}
\end{figure*}

\begin{figure*}
\begin{center}
\resizebox{0.40\hsize}{!}{\includegraphics*{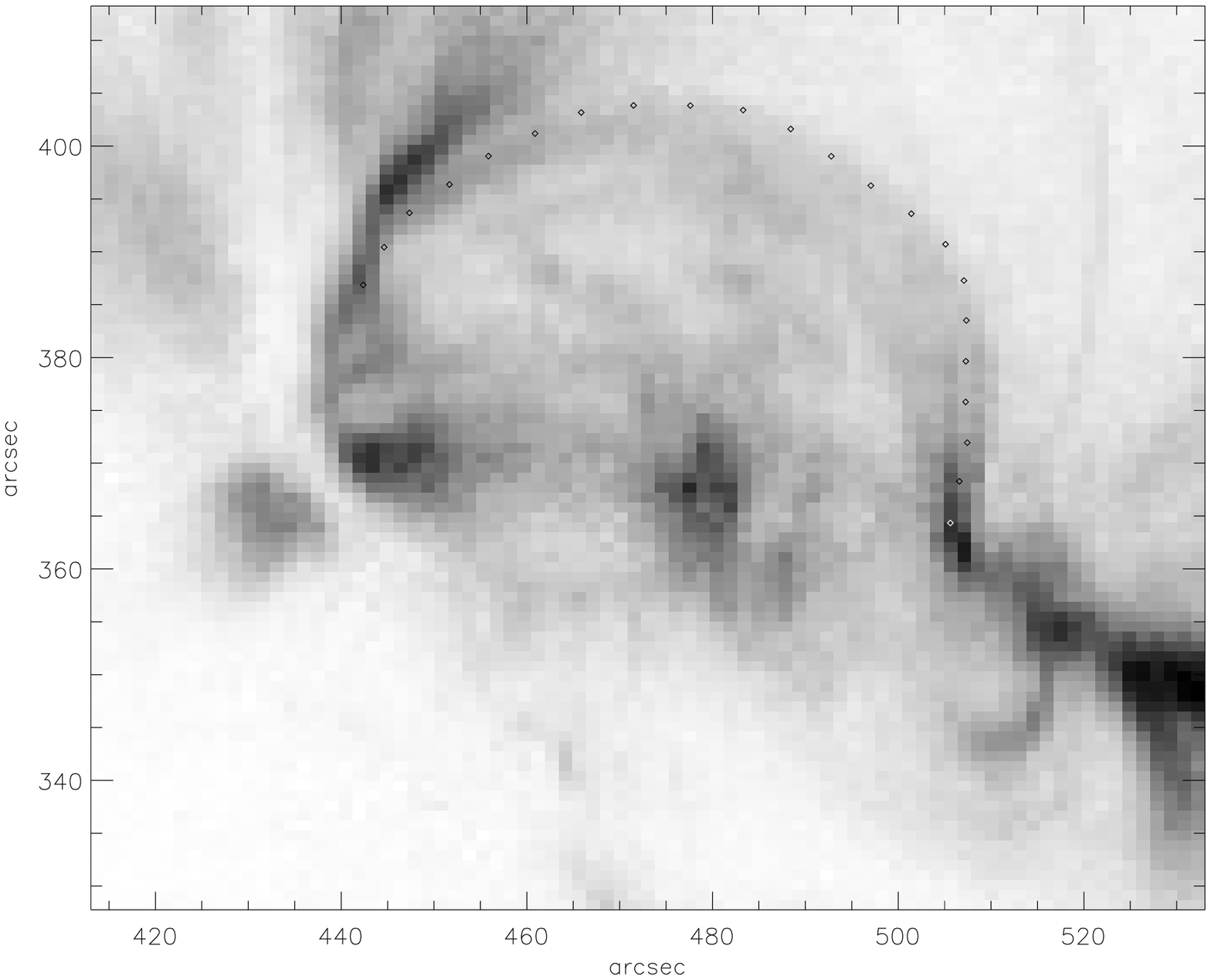}}
\resizebox{0.40\hsize}{!}{\includegraphics*{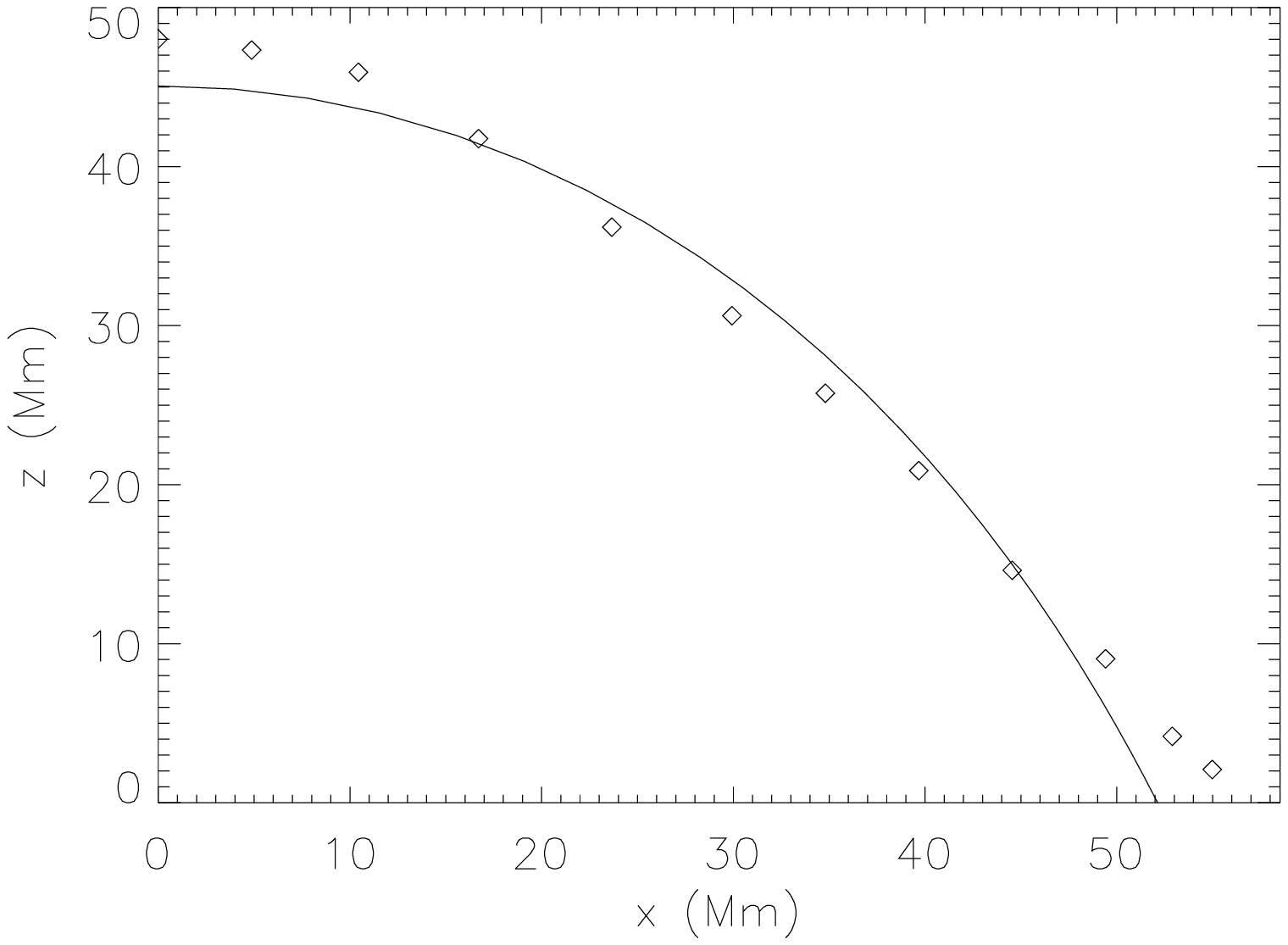}}
\end{center}
\caption{An MHD model of a loop observed by SUMER fitted to
observational data: shown are the SUMER image of the loop system
with the loop of interest indicated by a dotted line left
picture), and the model field line (solid line) fitted to the observed
line, represented by diamonds (right picture) in the $x$-$z$ plane.}
\label{sumerpics}
\end{figure*}

\begin{figure*}
\begin{center}
\resizebox{0.99\hsize}{!}{\includegraphics*{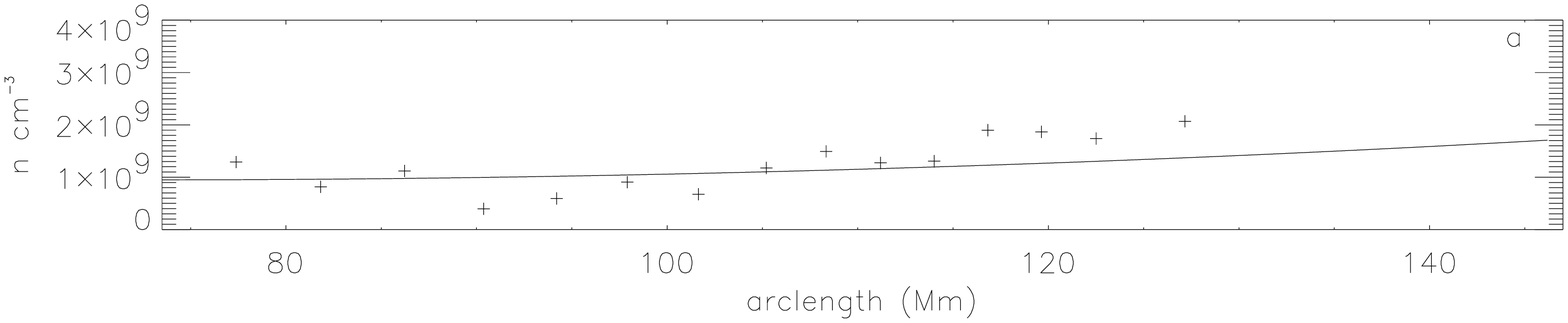}}
\resizebox{0.99\hsize}{!}{\includegraphics*{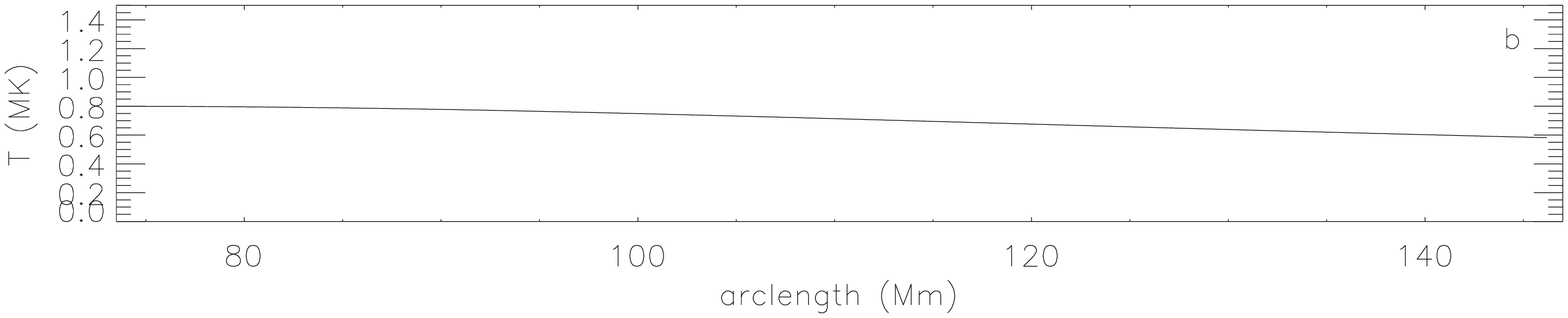}}
\resizebox{0.99\hsize}{!}{\includegraphics*{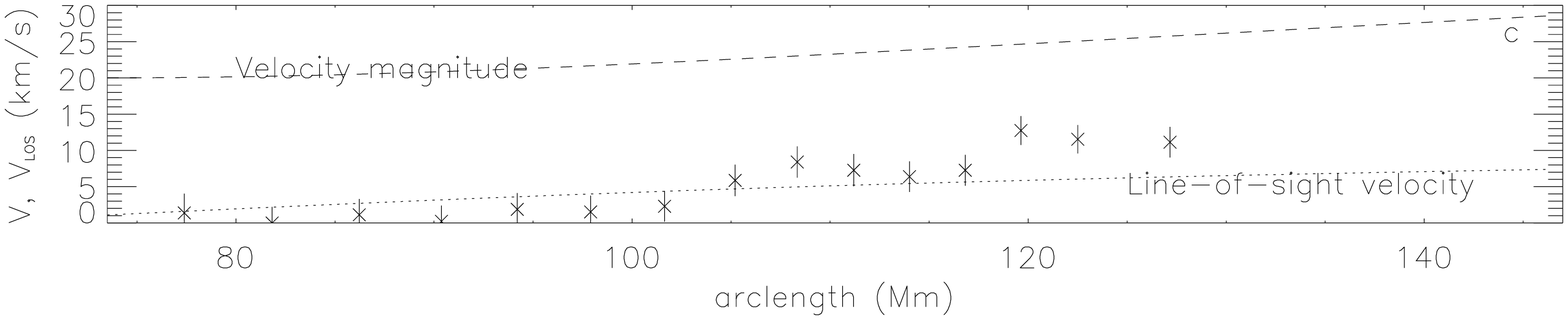}}
\resizebox{0.99\hsize}{!}{\includegraphics*{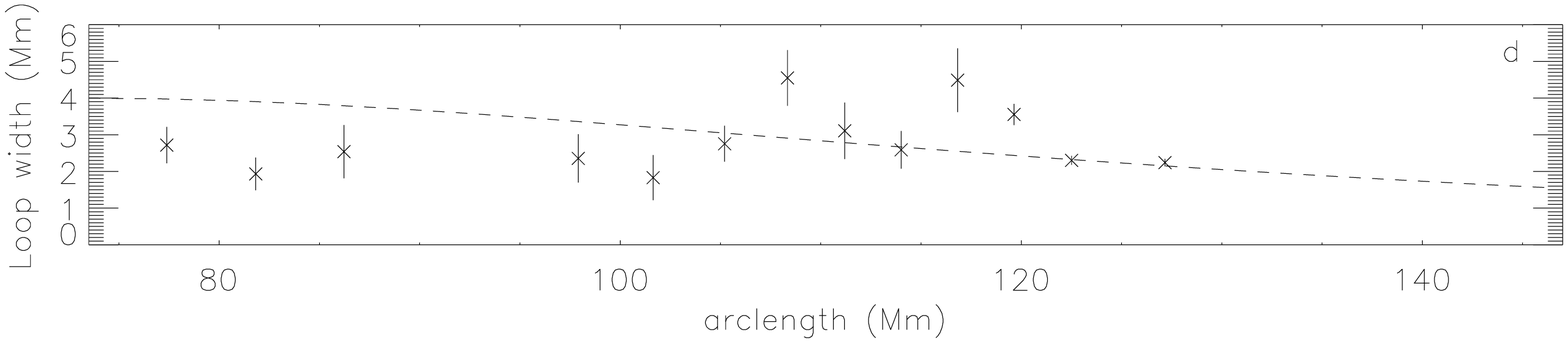}}
\resizebox{0.99\hsize}{!}{\includegraphics*{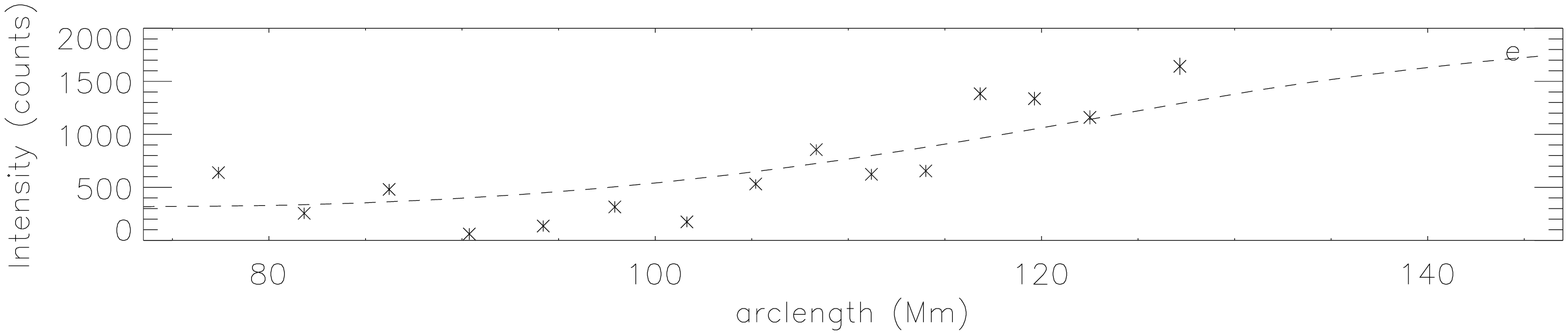}}
\end{center}
\caption{An MHD model of a loop observed by SUMER fitted to
observational data: the plots are organised as in Fig. \ref{loop2pics}.}
\label{sumerparams}
\end{figure*}

Figs.~\ref{loop2pics}, \ref{schmelzpics} and \ref{sumerpics} show pictures of
the image containing each loop and plots of solution field lines fitted to the
observed loop shape in the plane of the loop.  Figs. \ref{loop2params},
\ref{schmelzparams} and  
\ref{sumerparams} show plots of the density, temperature, together with the
absolute and
line-of-sight velocities of the models 
and comparisons of forward-modelled synthetic emission patterns compared to
the observed emission. 
Where available the observed values are also plotted.  It must be noted that
there are ambiguities in some aspects of
the fitting of the models to the data.  Coronal magnetic field observations are
not sufficiently advanced at present for a detailed model fit and so we impose
typical coronal field strengths in our models of $2$-$5$~G at the apex to $50$
- $100$~G, depending on field line geometry and inclination.  Although the
ODE's Eqs. (\ref{Gd}-\ref{P2}) depend on the magnetic field strength $|{\bf
B}|$ 
via $M$ we find that the value of $|{\bf B}|$ does not significantly affect the 
physical properties of the fluid.  We integrated Eqs.(\ref{Gd}-\ref{P2}) with
various 
start values of $|{\bf B}|$ up to a factor of $10$ greater and smaller than
those in 
the examples presented, keeping the start values of the other parameters 
fixed.  The only parts of the model changing significantly are the magnetic 
forces themselves in Figs.~\ref{loop2momen}, \ref{schmelzmomen} and
\ref{sumermomen} pictures a-d, while the other plots 
change very little.  An exception to this rule is the case where the magnetic 
field is too weak for the magnetic forces to be able to balance the other 
forces as seen in these pictures, in which case the integration simply 
fails indicating that an equilibrium is not possible.  
The effect on the system of varying $|{\bf B}|$ can be seen explicitly in
Eqs.~(\ref{Gd}-\ref{P2}).  The strong 
coronal magnetic field combined with the slow flow velocities observed in the 
corona together cause the flow to be very sub-Alfvenic ($M<<1$).  Hence varying 
$|{\bf B}|$ by a factor of $10$ has little effect on the size of $M$ compared
to the 
other variables, whose sizes are fixed by the observations.  It is for this
reason that the 
response of the plasma parameters to such variations in $|{\bf B}|$ is small.

There is some
ambiguity in the fitting
of the
temperature, density and velocity models, as well as the widths, due both to
difficulties in measuring
quantities along entire loop lengths precisely and to limits in the versatility
of the
solutions.  In the TRACE example Figs.~\ref{loop2pics}, \ref{loop2params}
observations of the 171~\AA\ and 195~\AA\ emission and filter ratio
calculations of the density and temperature are available along about half of
the loop length.  Elsewhere the emission is mixed with that of neighbouring
loops and so reliable measurements are not possible.  A measure of the shape
of the entire loop is available (see Fig. \ref{loop2pics}).  The filter ratio
measurements describe a near-isothermal loop whose density decreases with
height.  We are able to fit the density and temperature and emission patterns
of this loop reasonably for much of the region where observations are
available.  In the CDS example, the DEM temperature and density measurements
from Schmelz et al.~(2001) are multi-thermal while our emission data are
extracted from a single Mg~{\sc IX} image.  Since the emission data and the
density and temperature data are inconsistent with each other our approach is
to concentrate on forward-fitting our MHD model to the emission data while
using the density and temperature data as a guide, providing information about
the density distribution and the temperature magnitude.  On fitting the
emission model to the observations we find a reasonable fit to the DEM density
data, and we find a temperature model which is within the range of the
temperature data although more nearly isothermal.  More isothermal models or
models with flatter density profiles could not be made to fit the observed
emission.  A fit of T closer to many of the Schmelz et al.~(2001) temperature
data points at around 2~MK in this example requires a density function larger 
than the Schmelz et al.~(2001) density data, which seems very unlikely since
many loops seem to 
contribute to these density measurements.  Besides, the Mg~{\sc IX} line at
1~MK is significantly cooler than 2~MK and so we would expect this image to
pick up some of the cooler strands of this loop system.  The velocity
measurements derive
from a dopplergram from this same Mg~{\sc IX} image and, taking the angles of
the loop geometry and tilt into account as described in
Sect.~\ref{construction}, we are able to model these measurements to
reasonable
accuracy.  Filter ratio or DEM measurements of the density and temperature for
the SUMER example are not possible, and so density measurements are calculated
from a single Ne~{\sc VIII} image using the response function and taking the
temperature to be $0.7$~MK.  Velocity measurements are also extracted from
this same image.  While these measurements are more scattered than those of
the TRACE and CDS examples, approximate fits of the MHD model to the
intensity, density and velocity measurements with a near-isothermal
temperature model at around $0.7$~MK are given.

A measure of the width of the loop is made difficult by the mixing of emission
with
neighbouring loops in all three examples and low resolution of the
instruments in the CDS and SUMER examples.  Therefore there is much
uncertainty in these measurements.  Furthermore, because of the self-similar
structure
of the solution class (see Sect.~\ref{analyticalmodel}) the profile of the
width
of a model loop is defined by the shape of the loop so that a solution fitting
both the observed field line shape and the observed loop width is not
generally possible within our models.  The expanding cross-sections derive
directly from the self-similar structure of the solutions as described in
Sect.~\ref{construction} which for the moment we cannot avoid, since the
self-similar assumption embodied by Eqs. (\ref{M2},\ref{assumptions2}) is
crucial for us to solve the MHD equations.  The model widths are compared to
the observed widths in Figs.~\ref{loop2params}, \ref{schmelzparams} and
\ref{sumerparams}.

The flow in our
examples is unidirectional from one foot point to the other as in the models by
Cargill \& Priest~(1980, 1982) and Orlando et al. 
(1995a, 1995b). However, unlike those models the flow in our examples is not
sustained by a pressure difference between the loop foot points, the siphon 
mechanism.  This mechanism is included in the models by Cargill \&
Priest~(1980, 1982) and Orlando et al. 
(1995a, 1995b) because the only force which can
initiate in these models a unidirectional 
loop-aligned flow along the field lines is a suitable pressure gradient.  
However the physical details
of the initiation of flow in the corona are not well known.  The flow may not
have been initiated in a pre-existing loop, but may have been caused during
the loop's formation by the interaction of several forces.  Moreover in the
steady state such a pressure difference is not necessary to
maintain the flow.  A symmetric profile for the plasma inertia (signifying
e.g. acceleration up one loop leg and decceleraton down the other, or vice
versa)
may easily be balanced in a symmetric 
plasma model by gravity, the pressure gradient and, in two dimensions, by the
Lorentz force.  Because we are interested in modelling steady states, for
simplicity we choose to model symmetric loops which have pressure profiles
symmetric over the loop length.  Although the flow is unidirectional, we are
not describing siphon flows. 
We remark that the well-known "siphon flow" models of isolated flux tubes by
e.g. Thomas (1988) and Montesinos \& Thomas (1989) do not include pressure
differences despite their use of the term ``siphon flow''.

\subsection{Momentum balance}

\begin{figure*}
\begin{center}
\resizebox{0.99\hsize}{!}{\includegraphics*{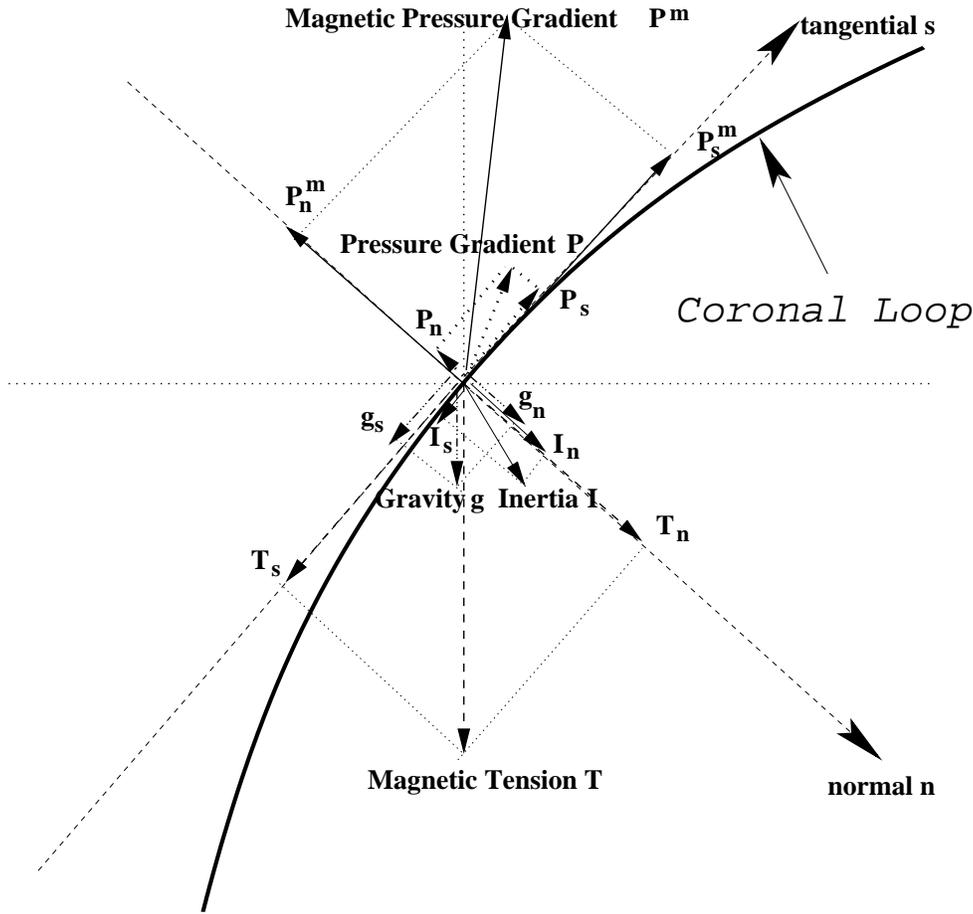}}
\end{center}
\caption{
The breakdown of momentum balance along and across a steady coronal loop. Shown
are the magnetic pressure gradient $P^m$, the magnetic tension force 
$T$, the
gas pressure $P$, the gravitational force $g$ and the inertial force $I$.
Also shown are the components of these forces resolved in the direction
tangent (normal) to the field, with subscript $s$ ($n$).
This diagram corresponds to the example momentum plots in 
Figs.~\ref{loop2momen},
\ref{schmelzmomen} and \ref{sumermomen} as described in the text.}
\label{loopforceanalysis}
\end{figure*}

\begin{figure*}
\begin{center}
\resizebox{0.49\hsize}{!}{\includegraphics*{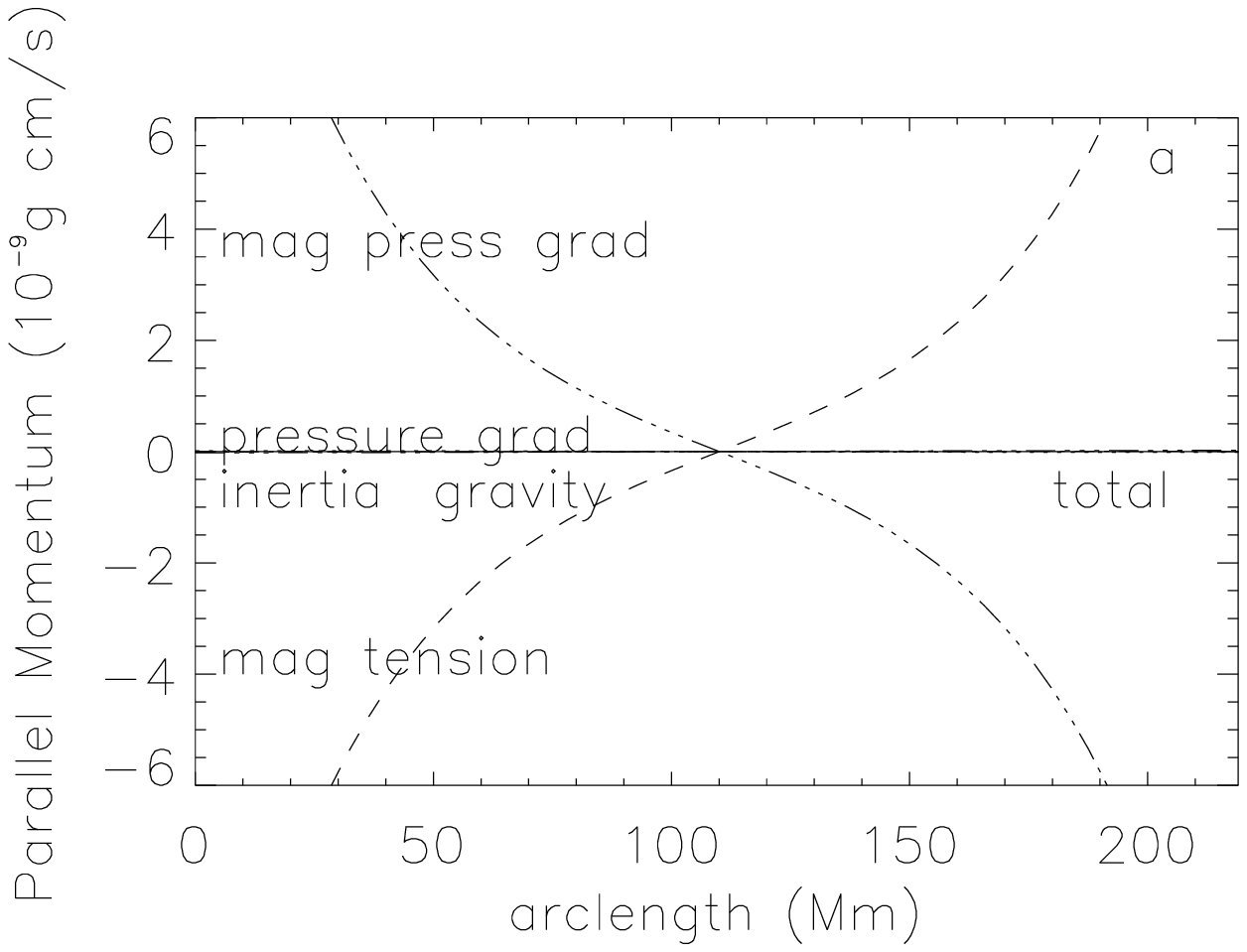}}
\resizebox{0.49\hsize}{!}{\includegraphics*{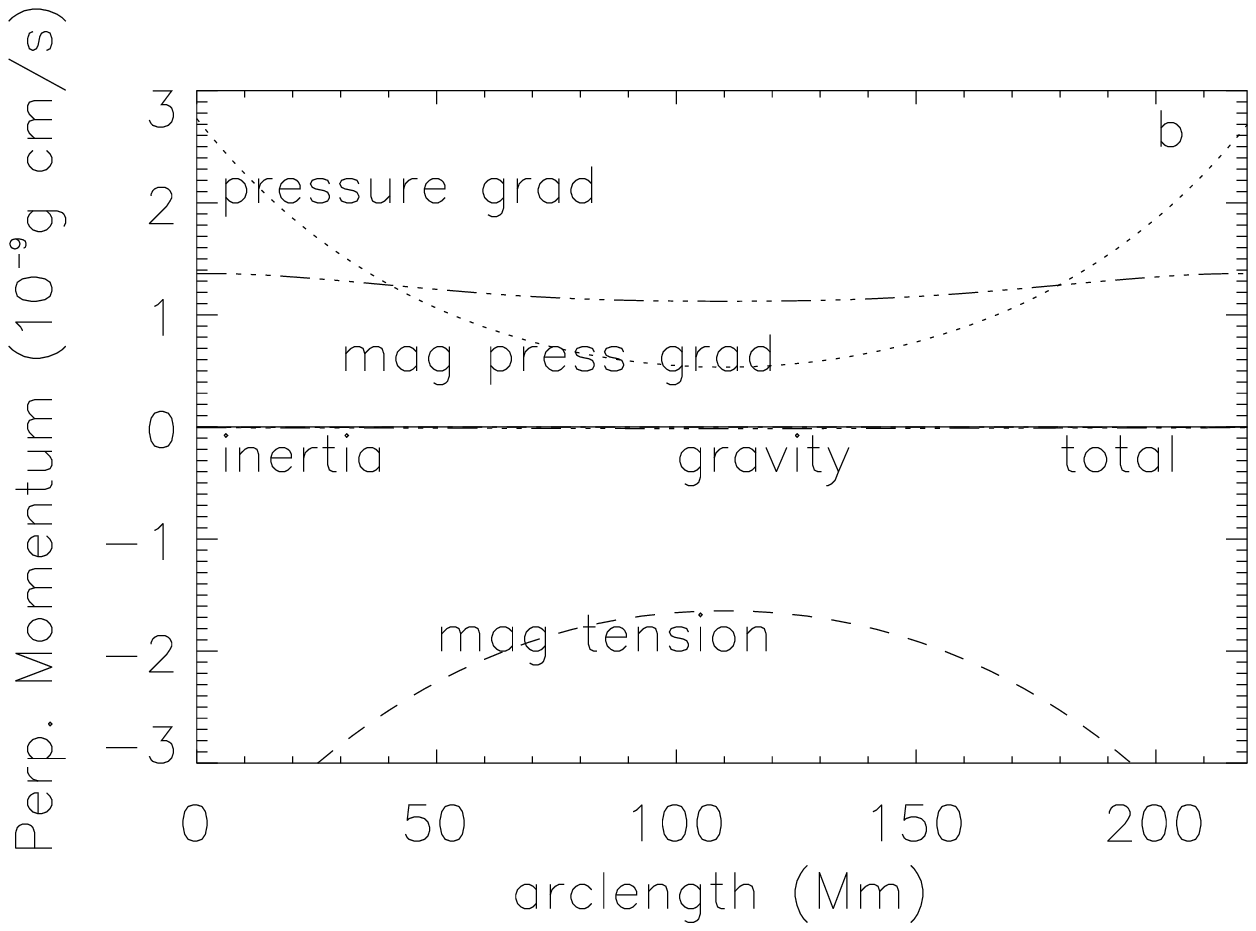}}
\resizebox{0.49\hsize}{!}{\includegraphics*{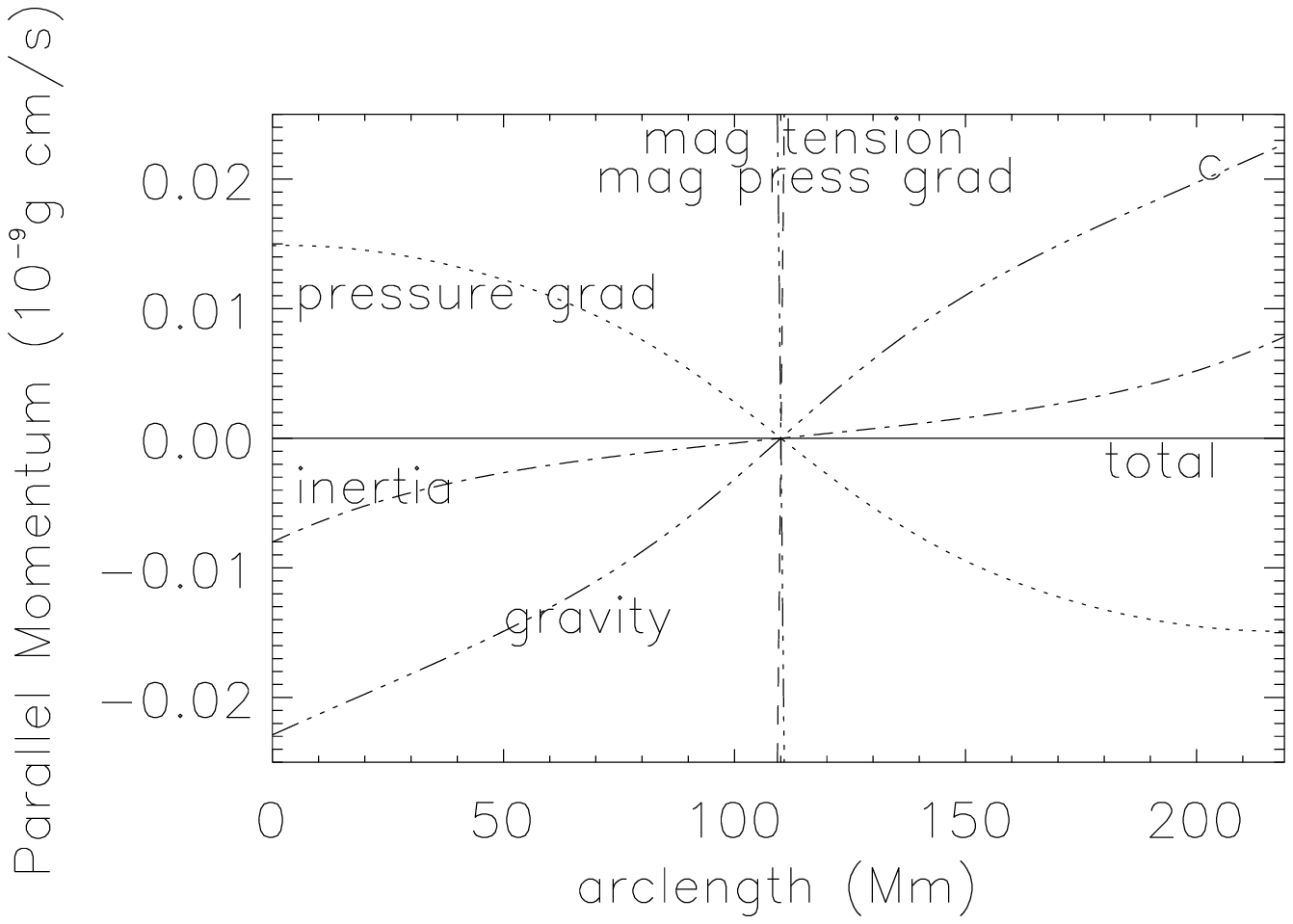}}
\resizebox{0.49\hsize}{!}{\includegraphics*{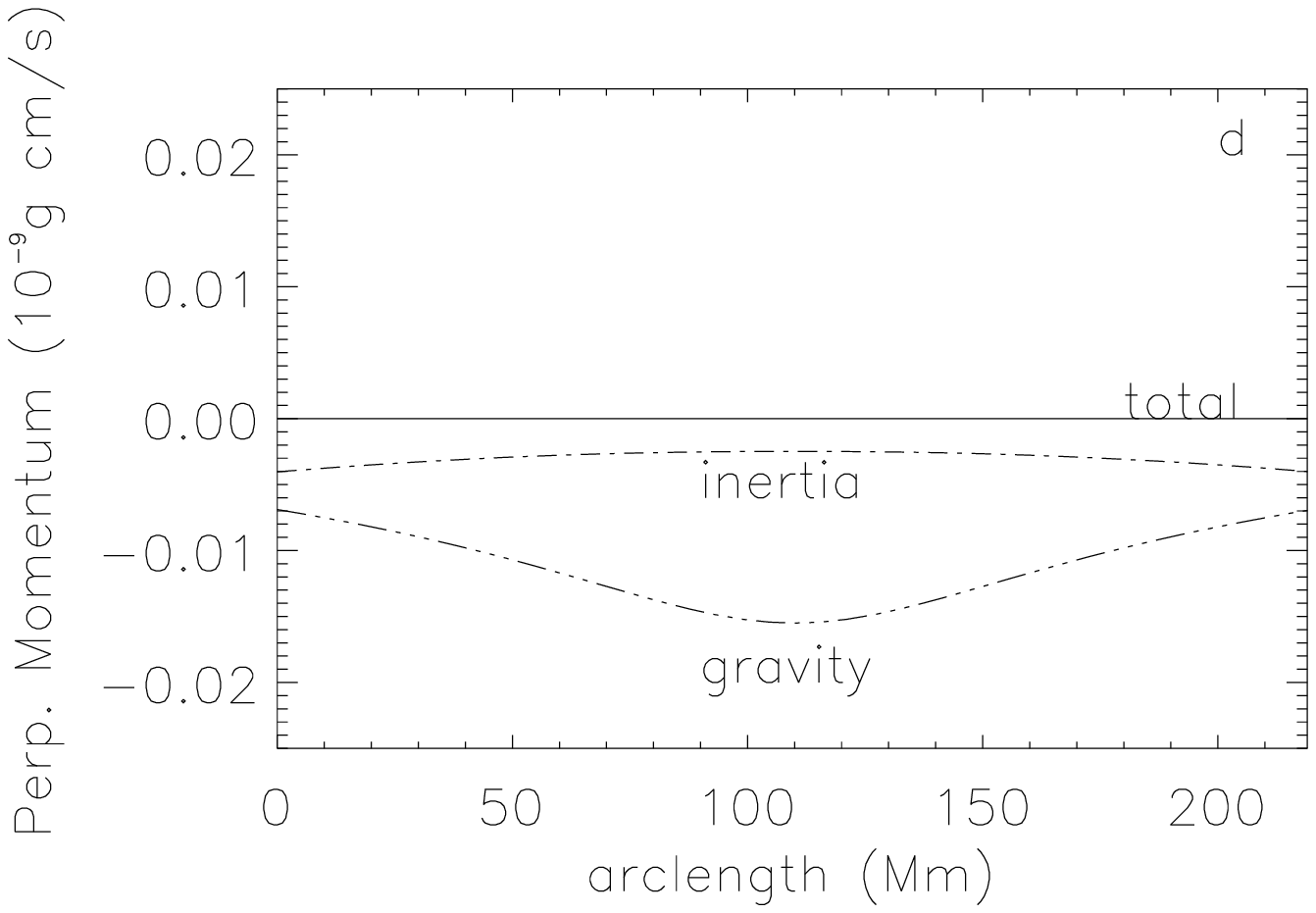}}
\resizebox{0.49\hsize}{!}{\includegraphics*{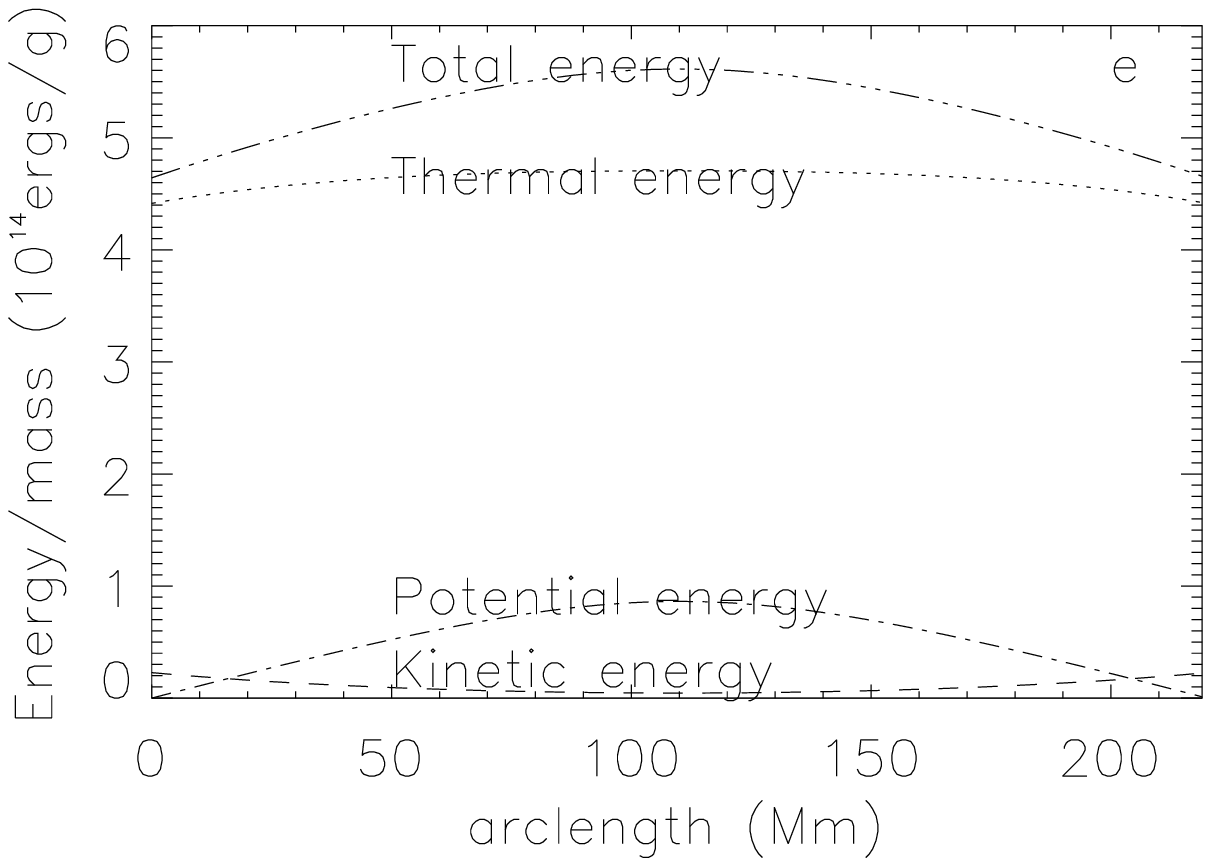}}
\resizebox{0.49\hsize}{!}{\includegraphics*{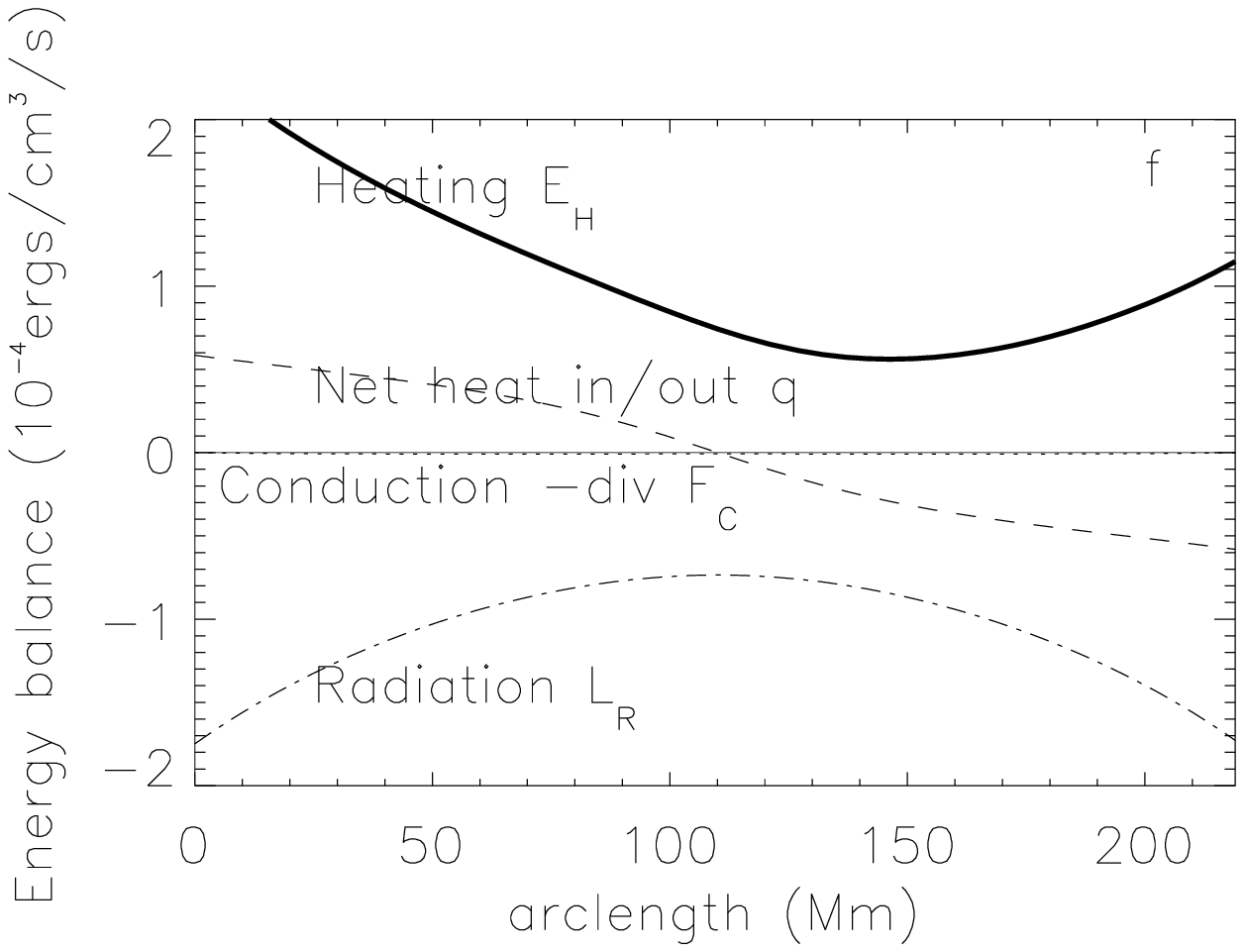}}
\end{center}
\caption{An MHD model of a loop observed by TRACE fitted to
observational data: shown are the breakdown of the momentum balance (a) along
and
(b) across the loop with magnified plots (c, d), (e) the energy integral along
the loop and (f) the volumetric energy
rate along the loop all graphed against arc length along the loop.  In the
momentum-along pictures, positive momentum means momentum directed from the
left foot point to the right, while in the momentum-across pictures, positive
momentum means momentum directed from inside the loop outwards.  In the
heating plot, the net heat in/out is represented by the dashed line,
the radiative losses by the dot-dashed line, the losses due to conduction by
the dotted line, and the remaining heating by the thick solid line.  Except
for a small region near the apex, radiative losses are
larger than conductive losses. The heating profile is largely
dominated by radiative losses but, influenced by the flow, it is
not symmetrical, but is concentrated at the inflow foot point.}
\label{loop2momen}
\end{figure*}

\begin{figure*}
\begin{center}
\resizebox{0.49\hsize}{!}{\includegraphics*{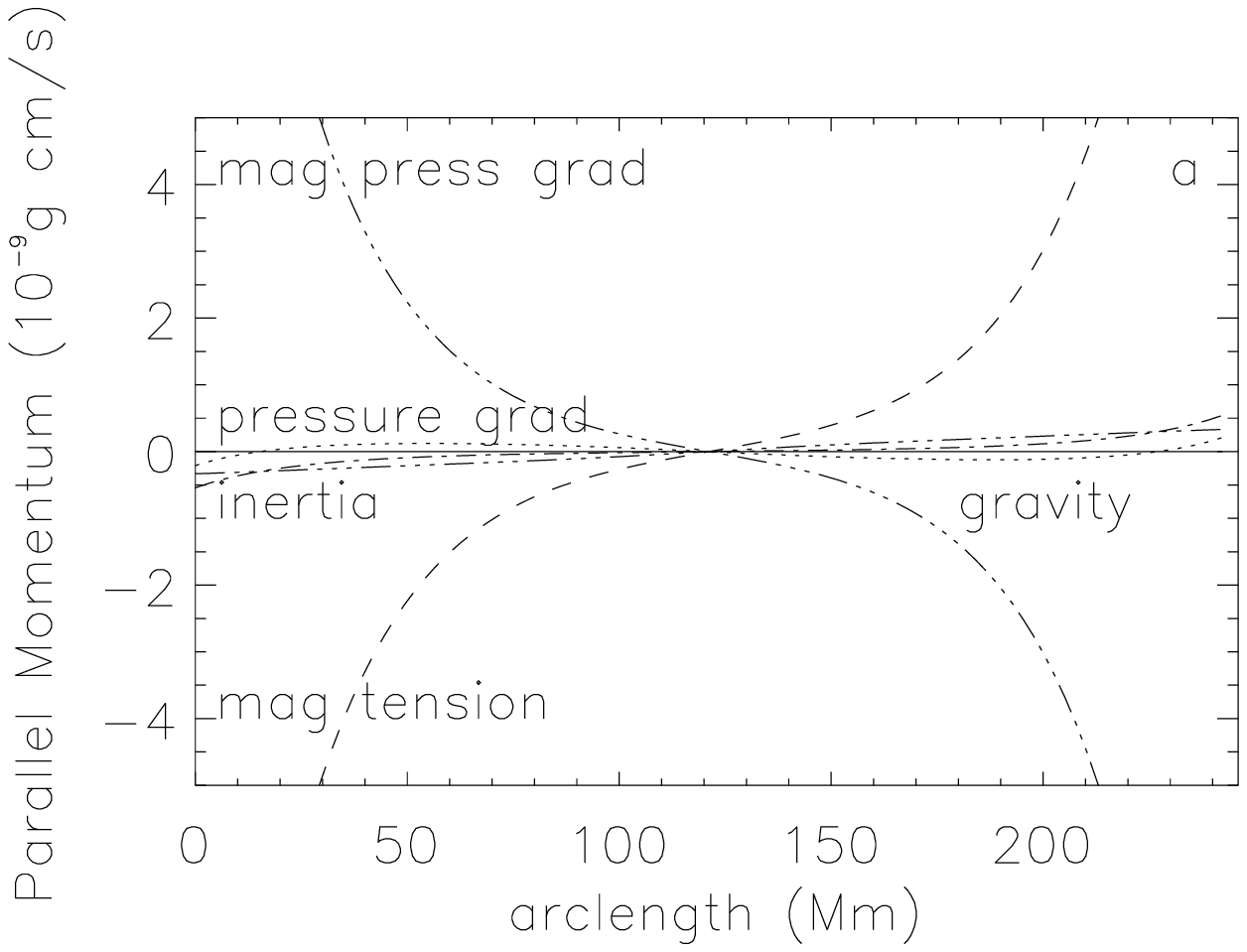}}
\resizebox{0.49\hsize}{!}{\includegraphics*{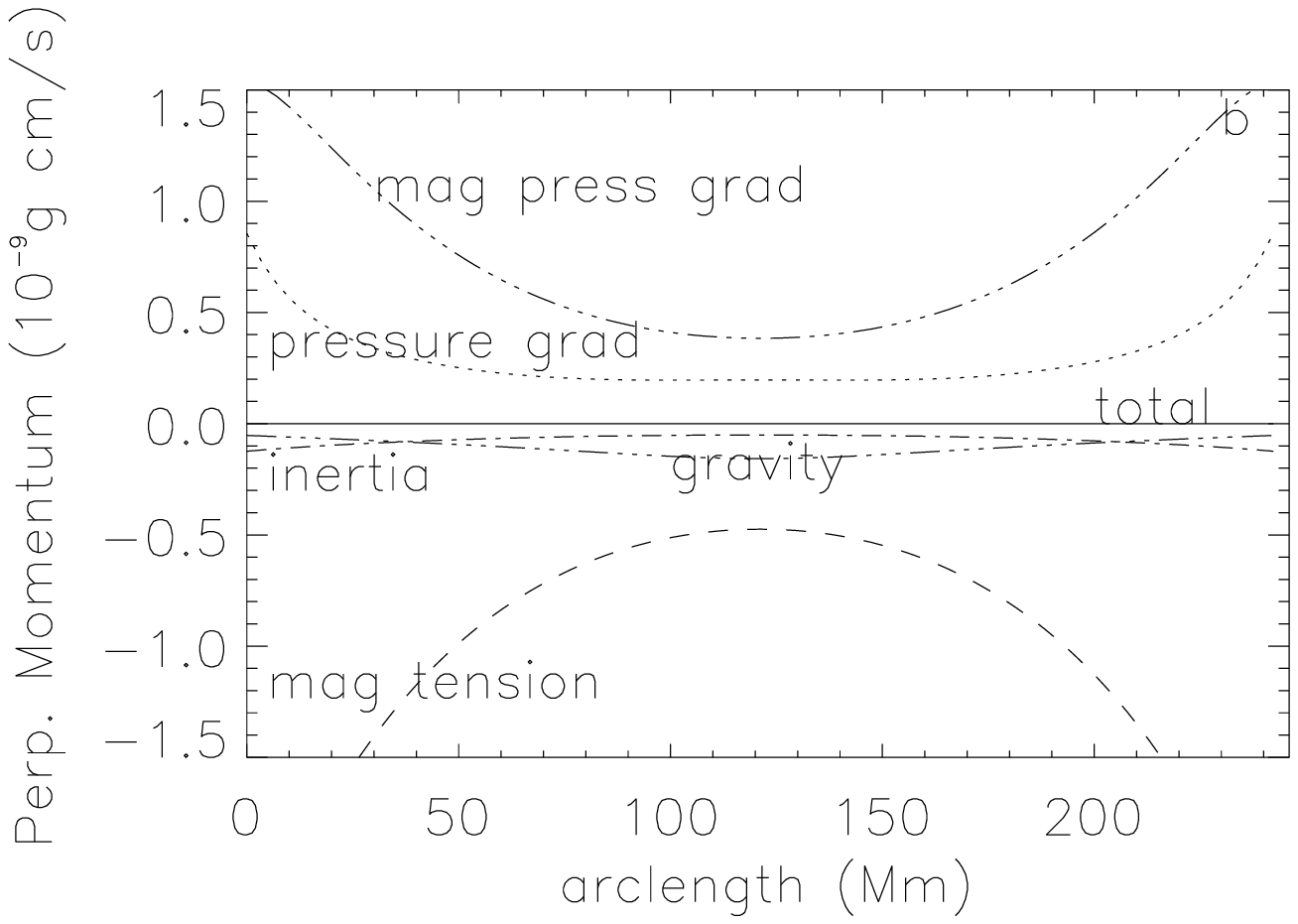}}
\resizebox{0.49\hsize}{!}{\includegraphics*{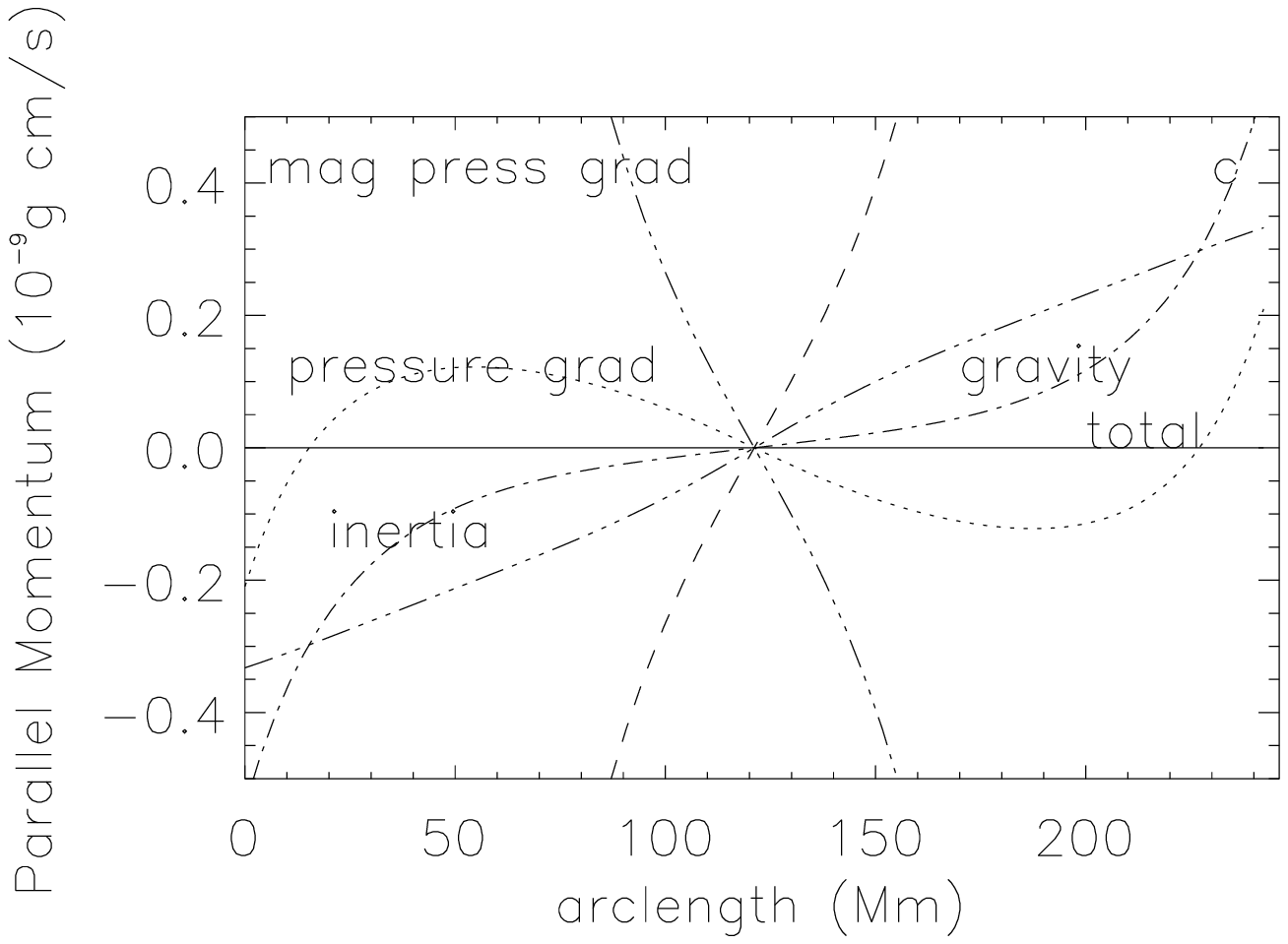}}
\resizebox{0.49\hsize}{!}{\includegraphics*{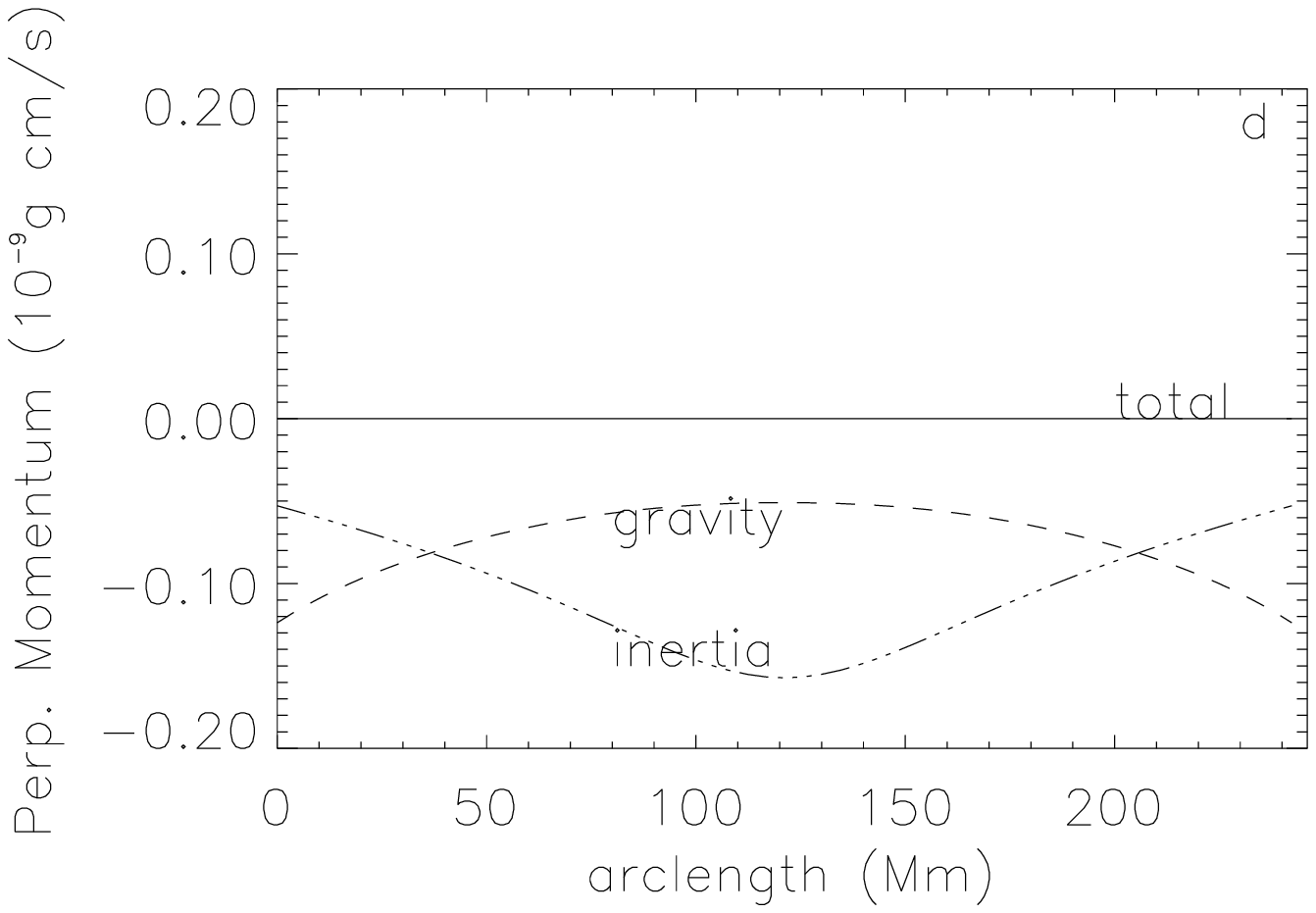}}
\resizebox{0.49\hsize}{!}{\includegraphics*{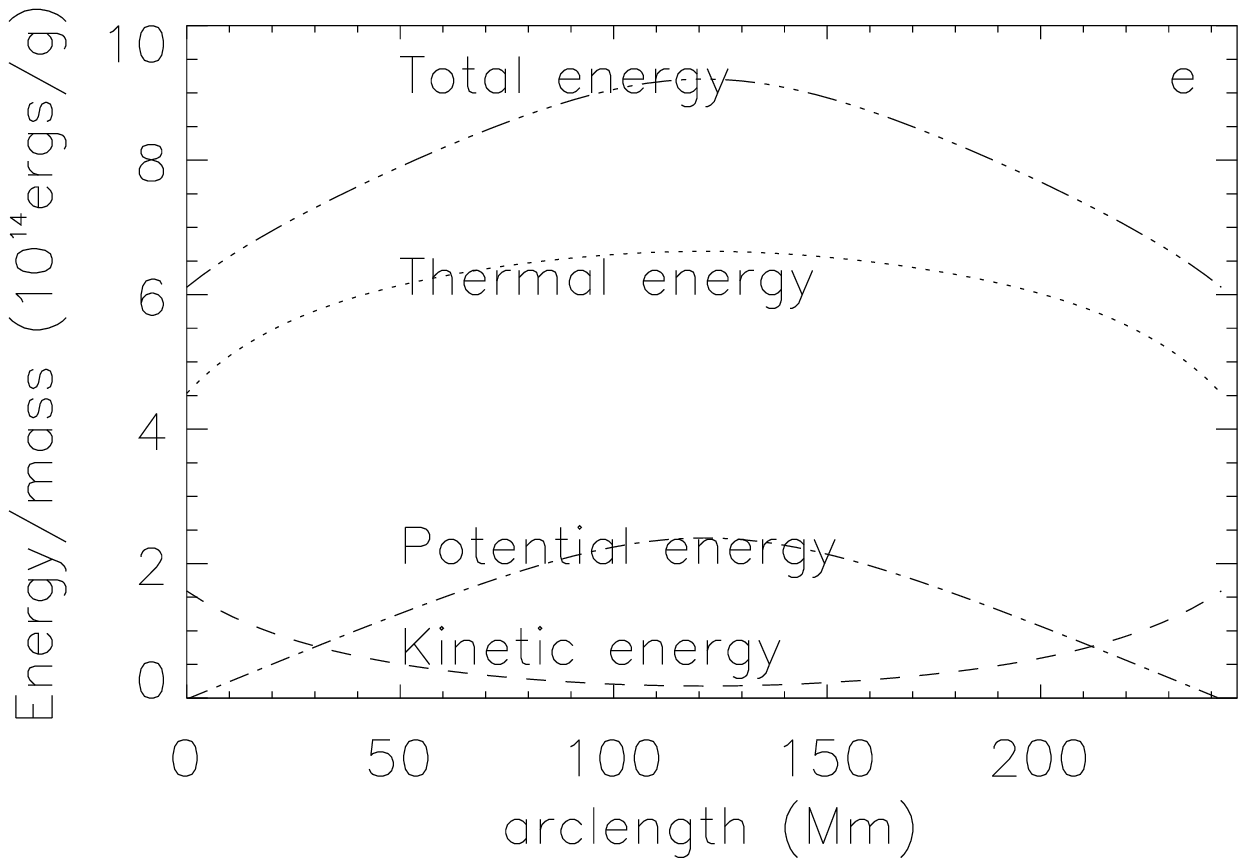}}
\resizebox{0.49\hsize}{!}{\includegraphics*{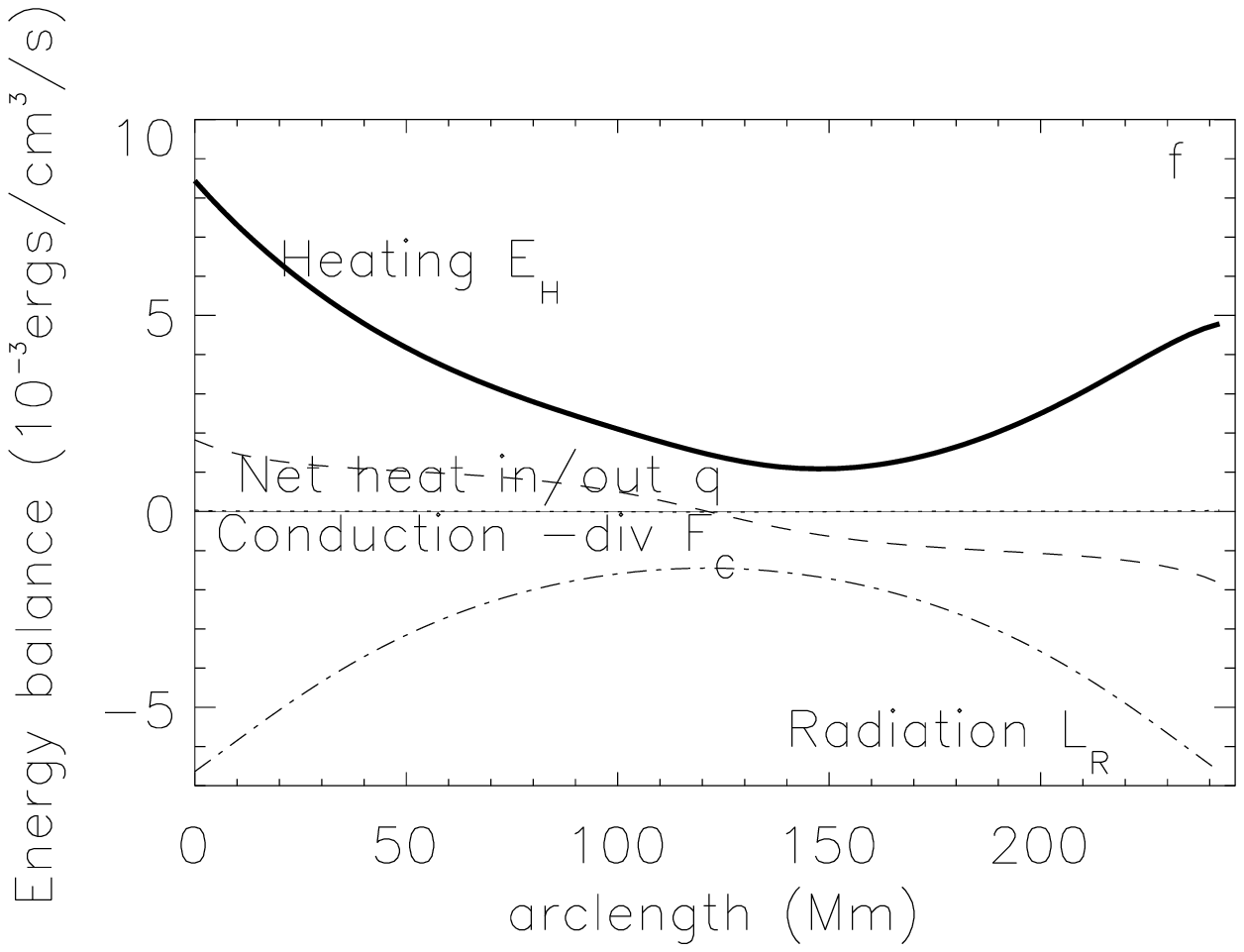}}
\end{center}
\caption{An MHD model of a loop observed with SoHO/CDS (Schmelz et al.,~2001),
fitted to
observational data: the plots are organised as in Fig.\ref{loop2momen}.  Note
that
here, as in
Fig.~\ref{loop2momen}, the radiative losses dominate the energy
balance, but an asymmetric heating function results because of the influence of
the flow.}
\label{schmelzmomen}
\end{figure*}

\begin{figure*}
\begin{center}
\resizebox{0.49\hsize}{!}{\includegraphics*{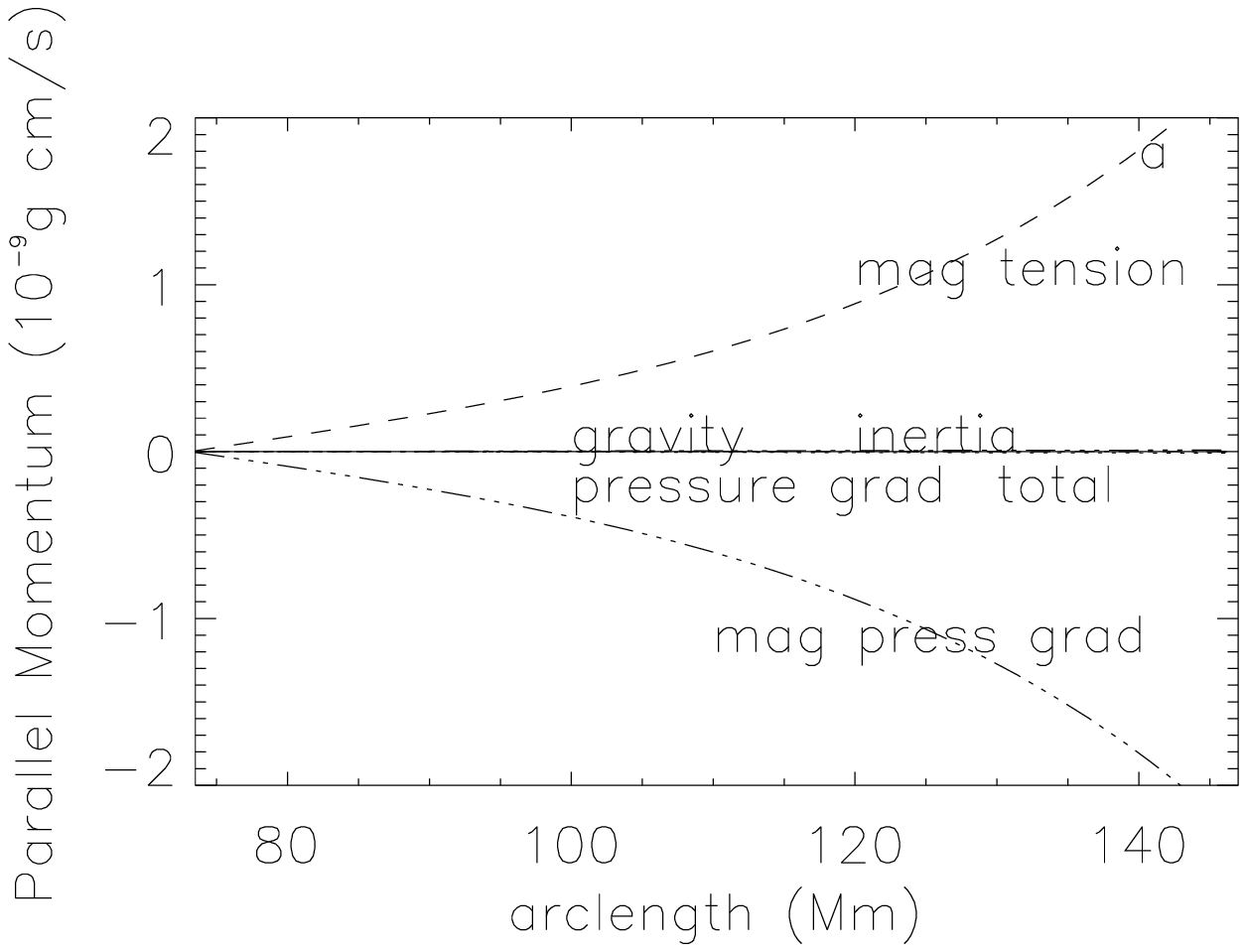}}
\resizebox{0.49\hsize}{!}{\includegraphics*{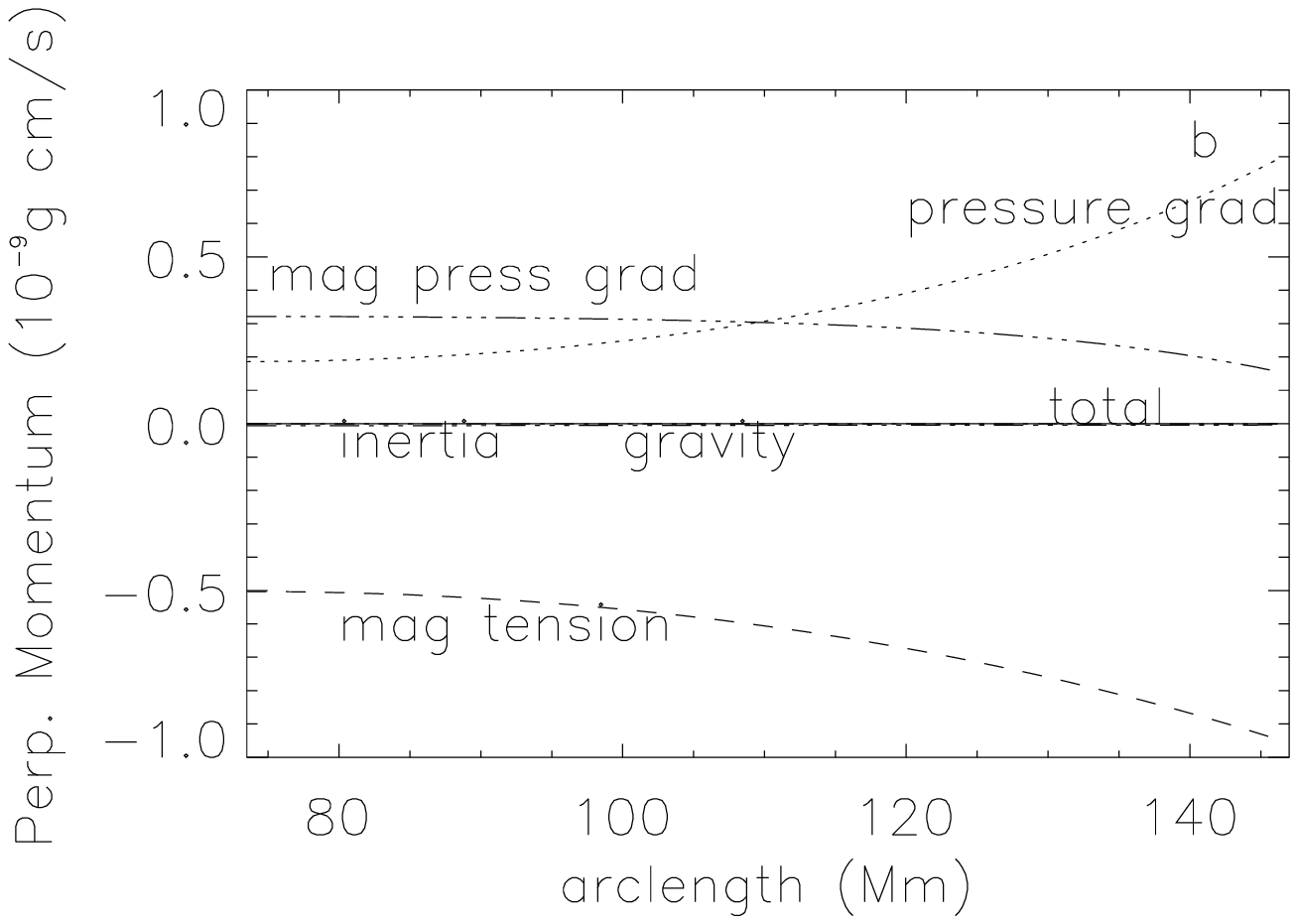}}
\resizebox{0.49\hsize}{!}{\includegraphics*{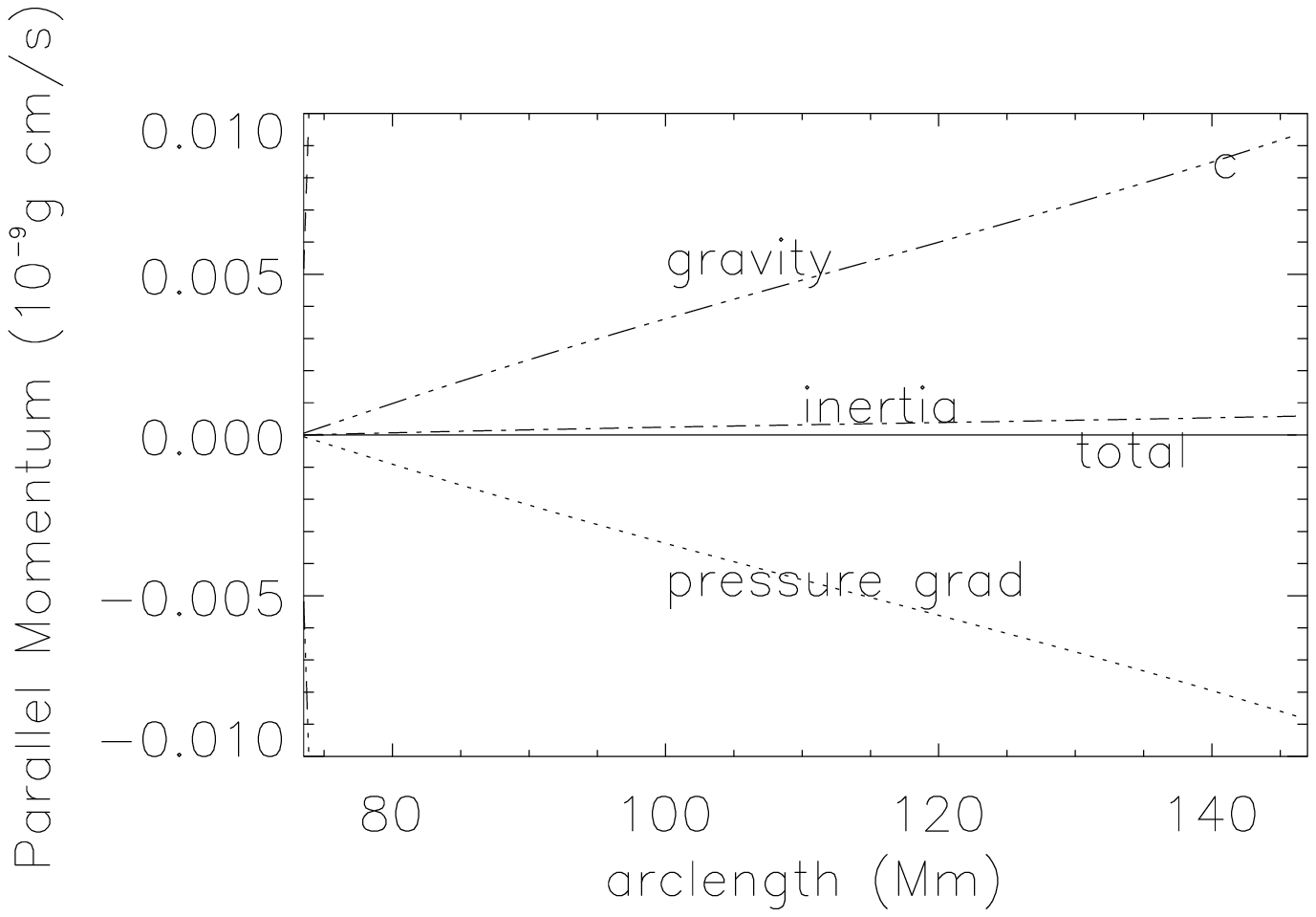}}
\resizebox{0.49\hsize}{!}{\includegraphics*{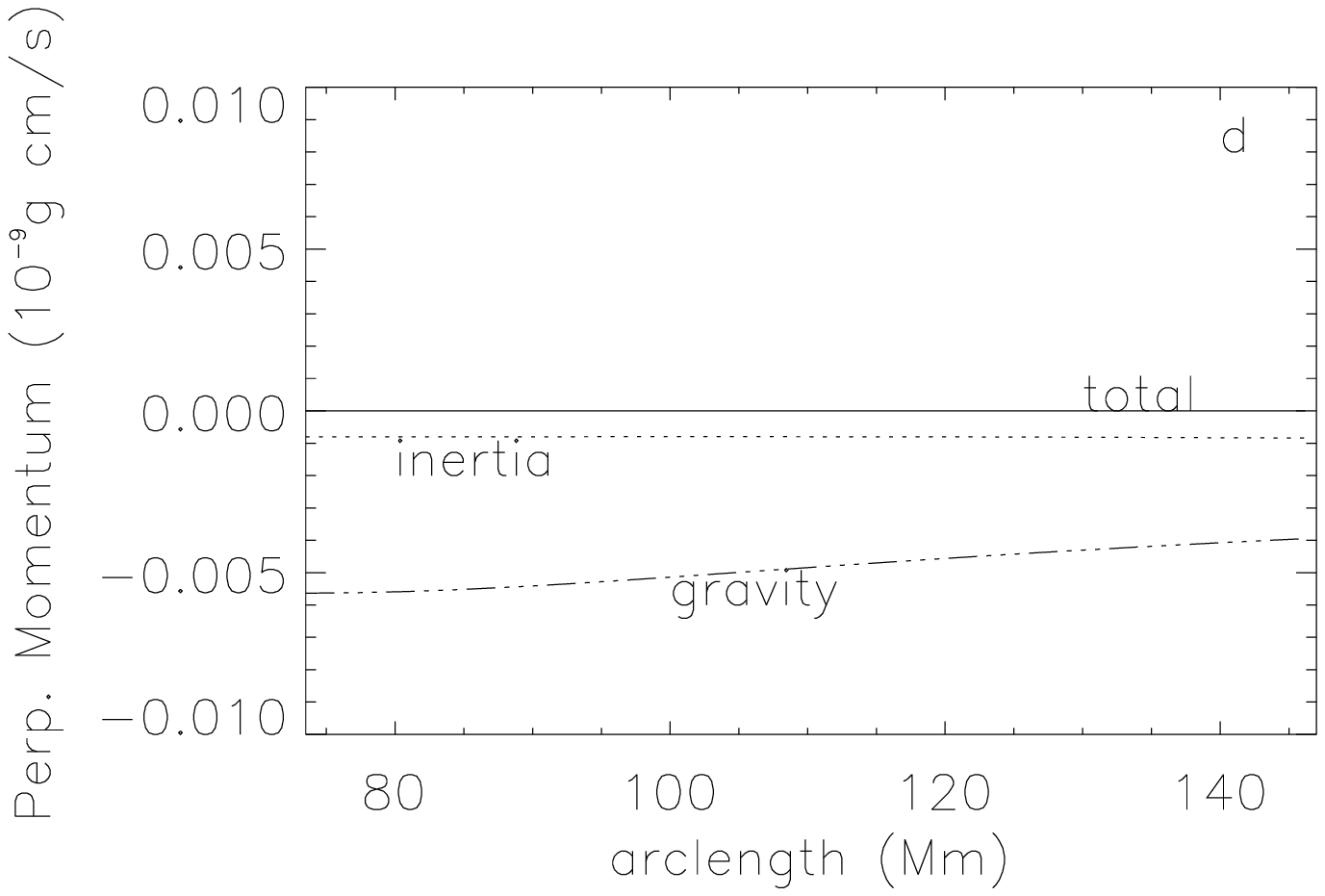}}
\resizebox{0.49\hsize}{!}{\includegraphics*{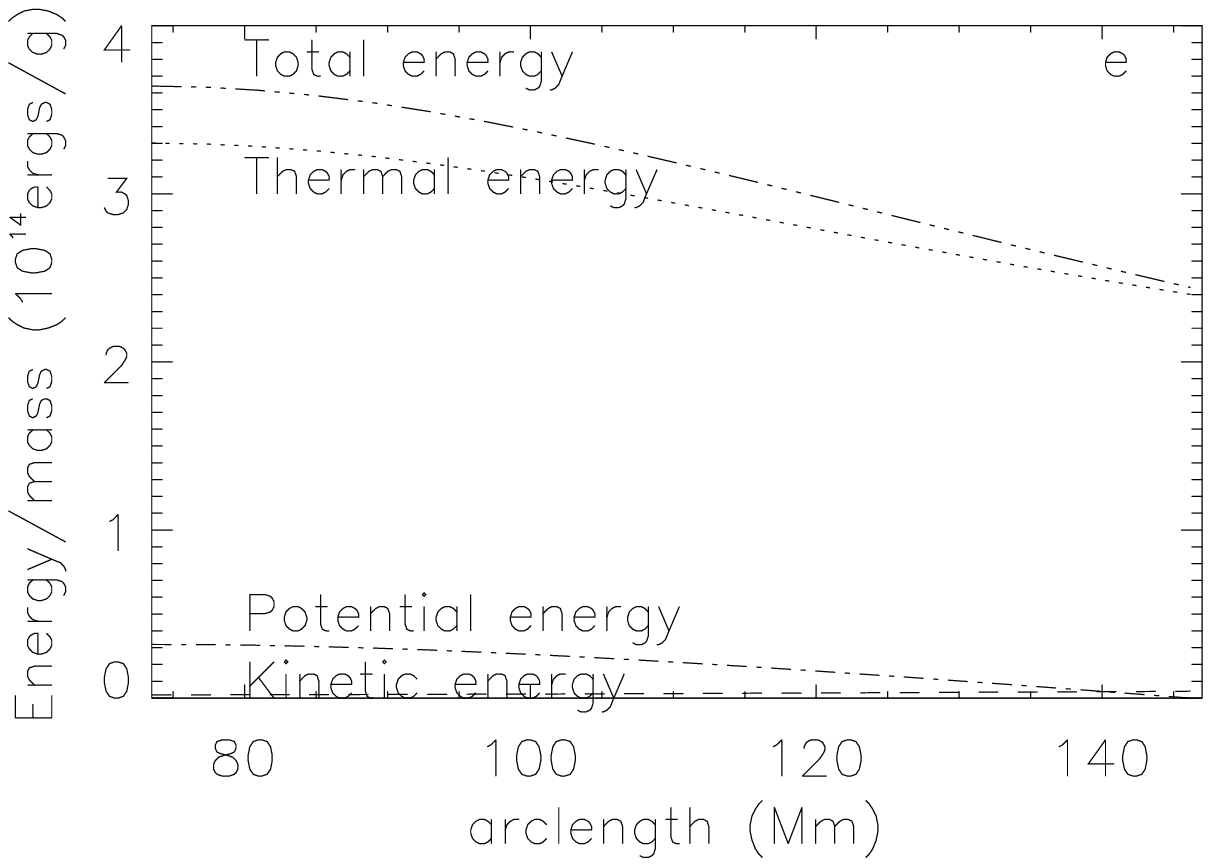}}
\resizebox{0.49\hsize}{!}{\includegraphics*{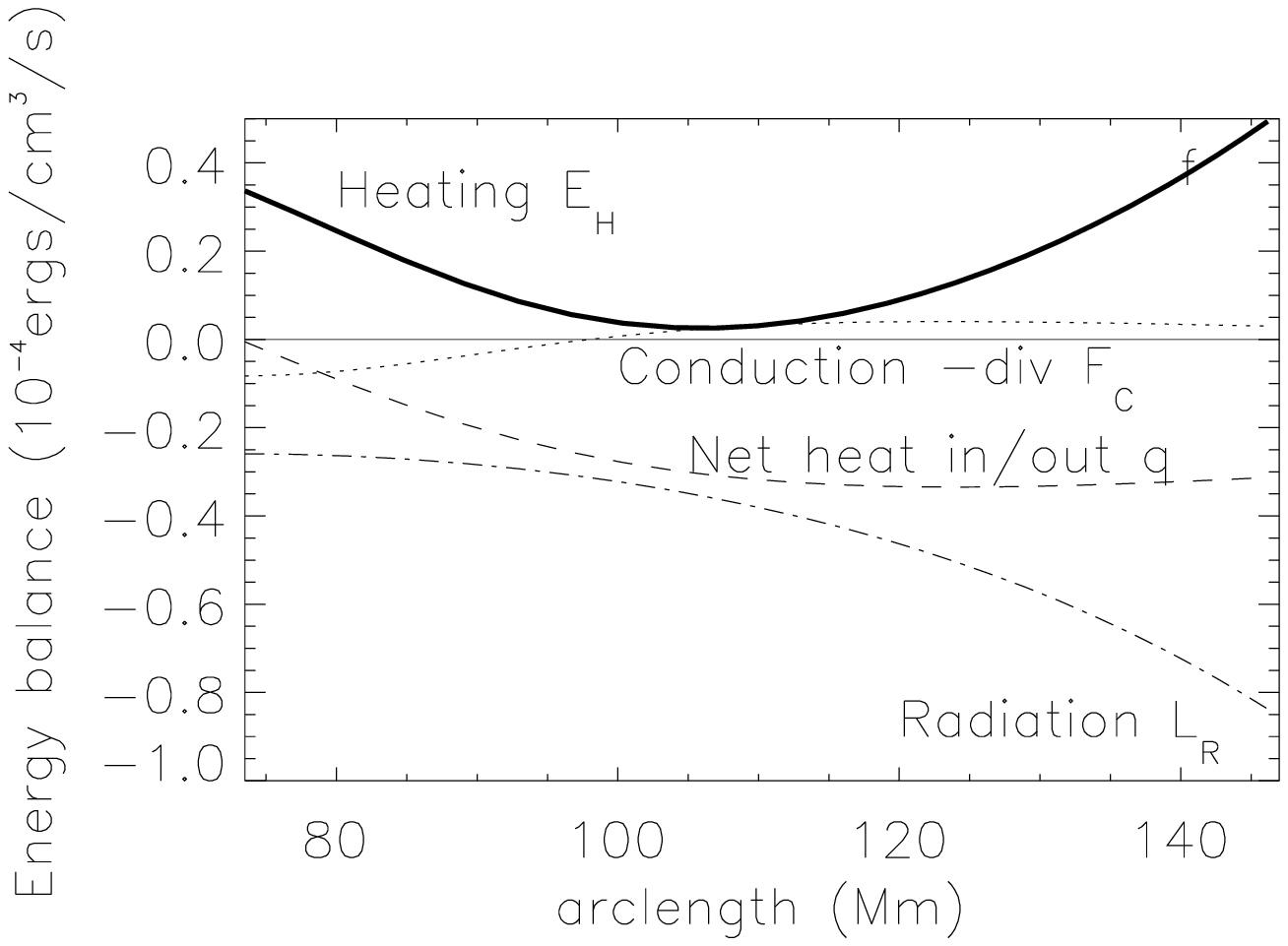}}
\end{center}
\caption{An MHD model of the loop observed by SUMER fitted to
observational data: the plots are organised as in Fig.~\ref{loop2momen}.  Note
that
in these plots the loop profiles resemble the profiles of the downflow (right)
leg of the TRACE and CDS examples in Figs.~\ref{loop2momen},
\ref{schmelzmomen}.}
\label{sumermomen}
\end{figure*}

Although coronal loops are well known to be magnetic structures, the component
of force balance along the loop excludes the Lorentz force, and so it has
become
common to model them as approximately one-dimensional structures imposing
hydrostatic (Rosner et al.,~1978; Serio et al.,~1981; Aschwanden et al.,~2001;
Aschwanden \& Schrijver,~2002) or steady hydrodynamic (Cargill \& Priest~1980,
1982; Orlando et al.,~1995a,b) equilibrium along the loop.  The inclusion
of a second cross-field dimension in our modelling allows the Lorentz force to
interact with the other forces across the loop and self-consistently to 
determine its shape and cross-section.  
Our models are the first loop models to include these cross-field effects fully 
and consistently.
Given the highly magnetised nature of the solar corona, inclusion of the
magnetic 
field is important in describing the loop dynamics.  Of particular importance
is the fact that on a curved loop in two dimensions the inertial term is not
field-aligned and so the plasma velocity may be greatly influenced by the
magnetic field as well as the other forces, unlike the one-dimensional case. 
Furthermore, in the siphon flow models of Cargill \& Priest~(1980, 1982) and
Orlando 
et al. (1995a, 1995b) the flow velocity is determined by the density for a
given loop
cross-sectional area which these authors impose as a free function, while in
our models the magnetic field selconsistently imposes the cross-sectional area
of the loop, thereby having a further direct influence on the plasma dynamics.  

Fig. \ref{loopforceanalysis} is an illustration of the breakdown of the 
momentum balance along and across a coronal loop into the five constituent 
forces: inertia of the plasma, magnetic tension, magnetic and gas pressure 
gradients and the gravitational force, as well as components of these forces 
resolved in directions tangent and normal to the field. This diagram
corresponds to the example momentum plots in Figs.~\ref{loop2momen},
\ref{schmelzmomen} and \ref{sumermomen} as described later in this subsection.
Note that the magnetic forces cancel in the tangential direction because 
the Lorentz force is  
perpendicular to the field. In the direction  normal to the field,  the 
Lorentz force
is non-zero and it is coupled with the remaining forces.  The magnetic tension
force acts vertically downwards because of the curvature of the loop.
The magnetic pressure gradient force has an upward vertical component 
because of
the vertically stratified structure of the magnetic field strength which 
decreases
with height. On the other hand, the magnetic field strength increases 
with slope  
because in an active region neighbouring field lines are generally bunched
close together near their foot points, generally located at a strong flux  
source/sink, and their separation increases with distance from the 
source/sink.
Hence, the horizontal magnetic pressure gradient force points towards the
interior of the loop.
The gas pressure gradient force has an upward vertical component because 
of the
stratification. On the other hand, the horizontal component of the gas 
pressure gradient force
points towards the center of the loop as the corresponding magnetic 
pressure gradient force does
because emission is generally found to be significantly higher in the 
region of active region loop
foot points than close to an apex;  in such near-isothermal structures, 
this implies that the 
gas pressure is higher at the loop foot points in comparison to the 
interior of the loop at the
same horizontal distance.
The normal component of the inertia points inside the loop towards the 
loop's centre
of curvature, as expected, while the tangential component is non-zero 
because the loop
is not circular and the curvature varies along the loop. In particular, 
it is negative
(i.e., it points towards the footpoints) because the curvature is 
increasing from
the left foot point to its maximum value at the apex.
The inertia on the right leg would be a mirror image of this, with a 
positive
tangential component indicating that the curvature is
decreasing away from the apex towards the right foot point.

Figs. \ref{loop2momen}, \ref{schmelzmomen} and \ref{sumermomen} show the 
breakdown of the momentum balance along the field and across the field, the 
volumetric energy profile along the loop and the volumetric energy rate 
per unit mass along the loop for the three models. 

As sketched in Fig.~\ref{loopforceanalysis}, in Figs.~\ref{loop2momen}, 
\ref{schmelzmomen} and \ref{sumermomen} the two components of the magnetic
force, 
the magnetic pressure gradient and tension oppose each other along and across 
each loop and they are significantly larger than the other forces acting along 
the loop, as is to be expected in a coronal model. 
The strength of the magnetic tension is greatest at the foot points, both along 
and across the field.  
In the CDS model of Fig.~\ref{schmelzmomen} the magnetic pressure gradient is 
maximum at the foot points.  This may be
surprising in the cross-field case, where a magnetic pressure gradient might
be expected to be almost field-aligned at a near-vertical foot point.  However
the pictures show that the increase in field strength towards the foot points
overcomes this effect to give a maximum magnetic pressure gradient across as
well as along the field at the foot points.  
In the SUMER example of \ref{sumermomen} the cross-field magnetic pressure
gradient decreases towards the foot points where the magnetic pressure
gradient is more nearly field-aligned than elsewhere.  
In the TRACE example shown in Figs. \ref{loop2momen}, the magmetic pressure
gradient 
is weaker than the gas pressure gradient everywhere across  the loop.  
Along the loop the magnetic forces cancel exactly since the Lorentz force must
be perpendicular to the loop.  Across the loop the magnetic forces are not
exactly balanced but they are again the largest forces.  
With a positive force in the cross field direction indicating a force away from 
the loop center of curvature, on the field-aligned plots in
Figs.~\ref{loop2momen}, 
\ref{schmelzmomen} and \ref{sumermomen} the upward/downward forces are 
positive/negative on the left half of the loop and negative/positive on the
right half.  

In a one-dimensional stratified hydrostatic atmosphere the gas pressure
gradient 
would point vertically upwards and decrease with height.  Along and across a 
loop standing in such an atmosphere the gas pressure gradient would appear in
Figs.
\ref{loop2momen}, \ref{schmelzmomen} and \ref{sumermomen}'s
field-aligned pictures as an odd function of arc length about the apex with a 
magnitude increasing with distance from the apex, as it happens in our model
too.  
In the cross-field pictures it would be represented by a positive even
function.  
The location of the
maximum pressure gradient across the field would depend on both the pressure
scale height and the shape of the loop, since the size of this component at a
point on the loop depends on both the size of the total pressure gradient and
the slope of the loop at that point.  For example, a vertical loop foot point
would have no pressure gradient across it in a one-dimensional stratified
atmosphere even though the total pressure gradient may have its maximum there. 
Similar comments apply to the gravitational force.  The gravitational force
would behave as the pressure gradient but with the opposite sign as the two
forces would balance in the one-dimensional stratified hydrostatic case.  

Our 2-D MHD model represents a significant departure from this situation as it 
can be seen from Figs.~\ref{loop2momen}, \ref{schmelzmomen} and
\ref{sumermomen}.  
In the field-aligned pictures the two components of the magnetic force, exactly
balancing each other, do not interact with the other forces.  The pressure
gradient and gravitational forces are more or less as in the hydrostatic case,
with a small contribution from the inertial force completing the force
balance.  However in the cross-field pictures of Figs.~\ref{loop2momen}, 
\ref{schmelzmomen} and \ref{sumermomen} there are significant
differences from the one-dimensional hydrostatic case.  The influence of the
magnetic forces on the other forces can be clearly seen: the two components of
the 
magnetic forces are not balanced across the loop and the net magnetic force
across the 
loop is negative. For example close to the footpoints of the loop, in
Figs.~\ref{loop2momen} 
and \ref{sumermomen}  it is the gas pressure gradient that is balancing
magnetic 
tension and only in Fig.~\ref{schmelzmomen} magnetic pressure balances tension.   
The gas pressure gradient, the only non-magnetic force positive across the
loop,
is significantly larger than the gravitational and inertial forces, and the
shape of the gas pressure curve may look very different from the gravitational
force curve in the full MHD case.  An important difference in this model
compared to one-dimensional hydrodynamic models is apparent in the
distribution of inertia along and across the flow field line.  Along the field
the inertia is maximum near the foot points where the loop is straightest, and
is zero at the apex where it changes sign.  Across the field the inertia is
significant over most of the loop length and is larger than the field-aligned
component around the apex where the field is most curved.

\subsection{Energy and heating}

The energy profile along
the loop is dominated by the thermal energy or enthalpy.  There are smaller
contributions from the potential and kinetic energies.  The thermal energy is
directly proportional to the temperature of the loop, so that an isothermal
loop would have a flat thermal energy distribution and hence a flatter total
energy profile than an equivalent non-isothermal loop with temperature maximum
at the apex would have.  The potential energy is proportional to the loop
height as a function of arc length and so the total energy clearly depends on
the loop's shape.  Because of its small size, the kinetic energy has little
influence on the total energy curve.  Over most of the loop
length in each case the kinetic energy is insignificant.  While the kinetic
energy seems not
to play a major role in the momentum balance of the loop, the plot of the
volumetric
heating rate along the loop shows that the velocity can have an important
influence on the heating profile of the loop.  This may be surprising, but
dimensional analysis shows that it is likely to be possible.  

If we take typical coronal values for the number density $n_e = 5\times 10^8
\rm{cm}^{-3}$ (giving a typical density $\rho_c=4.0\times 10^{16}\rm{g\
cm}^{-3}$) and the temperature $T_c=10^6 \rm{K}$, and a conservative estimate
for the velocity $V_c=10^6 \rm{cm\ s}^{-1}$, and if we take as a length scale
the hydrostatic scale height $Z_c=6.0\times 10^9 \rm {cm}$ then we find that
the corresponding typical potential energy per unit mass is $gZ_c=1.65\times
10^{14} {\rm erg}/{\rm g}$, the kinetic energy per unit mass is
$V_c^2/2=5.0\times 10^{11} {\rm erg}/{\rm g}$ and the enthalpy per unit
mass is $h_c=4.13\times 10^{14} {\rm erg}/{\rm g}$.  Thus the kinetic
energy is not significant compared to the other energies.  Meanwhile the
radiative loss function is $L_r=6.62\times 10^{-6} {\rm erg\ cm}^{-3}{\rm
s}^{-1}$, the conduction is $\nabla\cdot{\bf F}_c=6.42\times 10^{-5} {\rm erg\
cm}^{-3}{\rm s}^{-1}$ and the volumetric net heat in/out is $q_c=1.10\times
10^{-5} {\rm
erg\ cm}^{-3}{\rm s}^{-1}$.  In fact heat conduction plays a much smaller role
in our models than these numbers indicate because our temperature models are
close to being isothermal.  Other deviations from the order-of-magnitude
calculations occur in our models for similar reasons, but the difference
between the roles of the flow in
the energy and heating profiles is clear in both modelling (compare relative
importance of the kinetic energy in Figs.~\ref{loop2momen}e,
\ref{schmelzmomen}e, \ref{sumermomen}e and the net heat in/out of flow $q$ in
Figs.~\ref{loop2momen}f, \ref{schmelzmomen}f, \ref{sumermomen}f) and
order-of-magnitude calculations.

Returning now to the models, the net heat in/out of the
loop, being the field-directed derivative of the total energy, is an odd
function which is positive on one half of the loop and negative on the other. 
In the TRACE example of Figs. \ref{loop2pics} and \ref{loop2momen} it is
comparable in size to the radiative loss function, the dominant part of
the energy rate balance over most of the length of the loop.  In the SUMER
example of Figs. \ref{sumerpics} and \ref{sumermomen} the net heat in/out
exceeds the radiation close to the footpoint.   The radiative
losses are symmetric and are concentrated near the footpoints where the
density is greatest.  Heat conduction is negative at the location of
temperature maxima and positive at the temperature minima.
For loops with
temperature maxima at the apex, the conduction is peaked there but is still
smaller than the minimum of the radiation.  The heating profile is mostly a
combination of the radiative losses and the net heat in/out of the flow.  The
asymmetry of the heating function shows the influence of the flow.  The flow's
effect on the heating function is to distribute the remaining heating function
towards the upflow foot points, not to alter the total heating across the loop
as a whole.  Because the net heat in/out of the flow is an odd function which
integrates to zero along the loop length, a static loop which is otherwise
identical to this one would have a symmetric heating function with the same
total heating.  Note, however, that a static solution is a degenerate subcase
of this solution class and that setting the variable $M$ to zero would remove
much of the freedom in the system of ODE's.  An absolutely static model of a
given loop is not generally possible although the velocity magnitude can be
varied.  This combination of asymmetric heating functions and symmetric
intensity profiles has already been seen in the numerical hydrodynamic studies
of Mariska \& Boris~(1983) and Reale et al.~(2000b).  In studies of impulsive
heating giving qualitative agreement with observed
brightness evolution in TRACE images, Reale et al.~(2000a, b) and Peres,~(2000)
find that heating one foot point causes the other to
brighten first because of plasma compression there, and they caution against
straightforward interpretation of the observations to infer the location of
heating.

\section{Conclusions}
\label{conclusions}

The use of a two-dimensional compressible equilibrium solution of the full
ideal steady MHD
equations with consistent heating model has presented us with an opportunity to
study the magnetic field's
influence on the plasma dynamics and energetics and the flow's influence on the
heating profile self-consistently for the
first time.  Previous loop models have been one-dimensional and have ignored
the influence of the Lorentz force on the dynamics and of the magnetic field
configuration on the loop cross-sectional width, resulting in hydrostatic or
hydrodynamic models where the loop cross-sectional width is a free function
imposed by the modeller.  We find through fully consistent modelling that the
magnetic field governs the width of a loop and that there is much interaction
between the Lorentz force and the plasma inertia across the loop, as well as
among the inertia and all other forces along and across the loop.  There is a
significant component of inertia across curved structures in two dimensions,
not taken into account in one-dimensional models, which has a bearing on the
velocity profile and therefore the heating function of a loop.  While the
velocity plays a minor role in the energy profile of each loop, as is to be
expected in such sub-Alfv\'enic flow models, the inclusion of such flows is
found to influence the heating functions of the loops significantly.  Where
equivalent static models would have symmetric heating functions dominated by
balancing radiative losses concentrated near the foot points, the inclusion of
even very sub-Alfv\'enic flows alters this picture by introducing an
anti-symmetric component to the heating profile, resulting in an
asymmetric heating function, biased towards the upflow foot point.  These are
the conclusions to be drawn for the observations studied and the solution
class used to study them.  We have tried to fit the models to the data sets as
far as possible but some ambiguity remains.  In particular, the accuracy of
the loop width fits is compromised by difficulties in measuring
quantities along entire loop length and by limits on the versatility of the
solutions whose structure imposes loop widths on the
models which may be incompatible with observations.  
Our cross-sectional width model is defined by 
the loop height and foot point separation and cannot be freely chosen.  This 
introduces significant uncertainty into the fit of the model width to the 
data and some of this uncertainty is passed on to other components of the 
model, qualifying some of the conclusions drawn.  The loop width profile 
affects the velocity $\bf V$ of the flow, the net heat in/out of the flow $q$
and the 
heat conduction $-\nabla\cdot {\bf F}_C$.  Compared to a loop with expanding
cross-sectional 
area as in our models, a loop with constant cross-section whose physical 
properties are otherwise the same would have smaller velocities close to the 
foot points.  This would carry over to the net heat in/out q so that the 
heating function's asymmetry would be reduced in a model with constant 
cross-section, by a factor of between $2$ and $6$ compared to our models.  The 
width affects the heat conduction as shown by Eq.~(\ref{conduction}), where the
second term 
describes the effect of expanding/converging field lines.  In models with 
maximum temperature at the apex such as ours, field lines which converge 
towards the foot points inhibit heat conduction from the apex to cooler 
regions lower down.  The non-constant cross-section changes the conduction 
function significantly compared to a model with constant cross-section, but 
since the conduction plays a small role in the heating model this difference 
does not change the heating function significantly.  It is not known if the 
observed loops have constant or expanding cross-sections.  Even if some or all 
have constant cross-sections, we have demonstrated that plasma flow can have 
a visible effect on the heating distribution.  This and other smaller
uncertainties do not affect the broad conclusions drawn from the models and
are a small price to pay for a full MHD treatment and the physical insight
that this affords.  In the future we intend to establish more general
patterns by modelling more data sets and by applying more solution classes
from Table 1 of Paper 1.

Of course some loops in the solar atmosphere are far from
equilibrium and the heating mechanism may be highly non-steady.  A full
time-dependent MHD treatment of the evolution of a coronal loop is not
possible at present due to theoretical difficulties.  In the meantime it is
important to clarify the more basic steady states.

Equilibrium solutions of the MHD equations have been applied in modelling
coronal and chromospheric structures in one or two dimensions in the past (see
Paper 1).  One criticism of such models is that,
although they model well the homogeneous macroscopic structure of the coronal
magnetic field, the corresponding homogeneity of the plasma parameters in
these models does not explain the well-defined and localised plasma emission
patterns familiar from observations\footnote{See Petrie \& Neukirch (1999) for
3D MHD flow equilibria where localising velocity and density along chosen
field lines is possible, although these solutions have other physical
disadvantages.}.  Models of flows in isolated magnetic flux tubes has
been carried out with full force-balance in
a hydrostatic medium by Thomas \& Montesinos (1990) and
Degenhardt (1989).  There has also been some effort to
model the effect of an external magnetic field on a magnetic
flux tube (with no flow) by balancing the magnetic tension of the tube against
buoyancy forces deriving from the ambient magnetic and gas pressures (Parker,
1981; Browning \& Priest, 1984, 1986).  However, in the corona plasma loop
structures trace out magnetic field lines whose
field strength and configuration are representative of the volume as a whole
(Bray et al.,~1991). 
Such flux tubes are referred to by Thomas~(1988) as ``embedded'' as opposed to
the ``isolated'' category of interest to these authors and their models cannot
describe the full equilibrium force balance for the corona.  It seems, then,
that full
equilibrium solutions of the MHD equations are the most appropriate approach
to modelling steady coronal structures.  Furthermore, judging from the
widespread application of global
equilibrium models, in particular the routine use of potential and linear
force-free field models, that this weakness is a small price
to pay for the benefits of equilibrium models.  We have demonstrated the
importance of a full treatment of the momentum balance in determining the
plasma dynamics and thermodynamics of a coronal loop.

\begin{acknowledgements}
GP and CG acknowledge funding by the EU Research Training Network PLATON, 
contract number HPRN-CT-2000-00153. CG also acknowledges support by the 
Research Committee of the Academy of Athens.
\end{acknowledgements}


\begin{thebibliography}{}

\bibitem[2001]{Aschwanden01} Aschwanden, M.J. 2001, ApJ 559, L174

\bibitem[2002]{Aschwanden02} Aschwanden, M.J. 2002, ApJ 580, L79

\bibitem[2003]{Aschwanden03} Aschwanden, M.J. 2003 ``Physics of the Solar
Corona'', Springer and Praxis, in preparation

\bibitem[1999]{Aschwandenetal99}
Aschwanden, M.J., Newmark, J.S., Delaboudiniere, J-P., Neupert,
W.M., Klimchuk, J.A., Gary, G.A., Portier-Fozzani, F. \& Zucker, A.
1999, ApJ 515, 842

\bibitem[2000]{AschwandenNitta} Aschwanden, M.J. \& Nitta, N. 2000, ApJ 535,
L59

\bibitem[2002]{AschwandenSchrijver}
Aschwanden, M.J. \& Schrijver, C.J. 2002, ApJS 142, 269

\bibitem[2001]{Aschwandenetal01}
Aschwanden, M.J., Schrijver, C.J. \& Alexander, D. 2001, ApJ 550, 1036

\bibitem[1991]{Bray} Bray R.J., Cram L.E., Durrant C.J. \& Loughhead R.E. 1991,
``Plasma Loops in the Solar Corona'', Cambridge University Press

\bibitem[1984]{BrowningPriest84} Browning, P.K. \& Priest, E.R. 1984, Solar
Phys. 92, 173

\bibitem[1986]{BrowningPriest86} Browning, P.K. \& Priest, E.R. 1986, Solar
Phys. 106, 335

\bibitem[1980]{Cargill1}
Cargill, P.J. \& Priest, E.R. 1980, Solar Phys. 65, 251

\bibitem[1982]{Cargill2}
Cargill, P.J. \& Priest, E.R. 1982, Geophys. Astrophys. Fluid Dyn. 20, 227

\bibitem[1994]{Cargill3}
Cargill, P.J. 1994, ApJ 422, 381

\bibitem[1994]{Cargill4}
Cargill, P.J. \& Klimchuk, J.A. 1997, ApJ 478, 799

\bibitem[2002]{Chaeetal} Chae, J., Park, Y-D., Moon, Y-J., Wang, H. Yun, H.S.
2002, ApJ 567, L159

\bibitem[1999]{Dammaschetal} Dammasch, I.E., Wilhelm, K., Curdt, W. \& Hassler,
D.M. 1999, A\&A 346, 285

\bibitem[2002]{Daraetal2002} Dara, H.C., Gontikakis, C., Zachariadis, Th.,
Tsiropoula, G., Alissandrakis, C.E., Vial, J.-C. 2002, in Proceedings of the
Second Solar Cycle and Space Weather Euroconference, ESA SP-477, 95

\bibitem[1989]{Degenhard89} Degenhard, D. 1989, A\&A 222, 297

\bibitem[1996]{delzannahood} Del Zanna L. \& Hood A.W. 1996, A\&A 309, 943

\bibitem[1990]{hoodanzer} Hood A.W. \& Anzer U. 1990, Solar Phys. 126, 117

\bibitem[1979]{HoodPriest} Hood, A.W. \& Priest, E.R. 1979, A\&A 77, 233

\bibitem[1999]{JudgeMcIntosh} Judge, P.G. \& McIntosh, S.W. 1999, Solar Phys.
190, 331

\bibitem[1957]{ks} Kippenhahn R. \& Schl\"uter A. 1957, Z. Astrophys. 43, 36

\bibitem[2000]{Mackay}
Mackay, D.H., Galsgaard, K., Priest, E.R. \& Foley, C.R. 2000, Solar Phys. 193,
93

\bibitem[1983]{MariskaBoris} Mariska, J.T. \& Boris, J.P. 1983, ApJ 267, 409

\bibitem[2002]{Martens} Martens, P.C.H., Cirtain, J.W. \& Schmelz, J.T. 2002,
ApJ 577, L115

\bibitem[1989]{MontesinosThomas89} Montesinos, B. \& Thomas, J.H. 1989, ApJ
337, 977

\bibitem[1981]{Parker81} Parker, E.N. 1981, ApJ 244, 631

\bibitem[2000]{Peres} Peres, G. 2000, Solar Phys. 193, 33

\bibitem[1999]{Peter} Peter, H. 1999, ApJ 516, 490

\bibitem[1999]{PeterJudge} Peter, H. \& Judge, P.G. 1999, ApJ 522, 1148

\bibitem[1999]{petrieneukirch}
Petrie, G.J.D. \& Neukirch, T. 1999, Geophys. Astrophys. Fluid Dyn. 91, 269

\bibitem[2002]{PVT01}
Petrie, G.J.D., Vlahakis, N. \& Tsinganos, K. 2002, A\&A 382,
1081 (Paper 1)

\bibitem[1995]{PorterKlimchuk}
Porter, L.J. \& Klimchuk, J.A. 1995, ApJ 454, 499

\bibitem[1982]{Priest}
Priest, E.R. 1982 {\it Solar Magnetohydrodynamics}, Reidel

\bibitem[1998]{Priestnat}
Priest, E.R., Foley, C.R., Heyvaerts, J., Arber, T.D., Culhane, J.L. \& Acton,
L.W. 1998, Nature 393, 545

\bibitem[[2000]{PriestApJ}
Priest, E.R., Foley, C.R., Heyvaerts, J. et al. 2000, ApJ 539, 1002

\bibitem[1977]{RaymondSmith} Raymond, J.C. \& Smith, B.W. 1977, ApJS 35, 419

\bibitem[2000]{Realeetala} Reale, F., Peres, G., Serio, S., DeLuca, E.E. \&
Golub, L. 2000a, ApJ 535, 412

\bibitem[2000]{Realeetalb} Reale, F., Peres, G., Serio, S., Betta, R.M.,
DeLuca, E.E. \& Golub, L. 2000b, ApJ 535, 423

\bibitem[1978]{RTV78}
Rosner, R., Tucker, W.H. \& Vaiana, G.S. 1978, ApJ 220, 643

\bibitem[1994]{st94} Sauty C. Tsinganos K. 1994, A\&A 287, 893

\bibitem[2001]{Schmelzetal2001}
Schmelz, J.T., Scopes, R.T., Cirtain, J.W., Winter, H.D. \&
Allen, J.D. 2001, ApJ 556, 896

\bibitem[2002]{Schmelz2002} Schmelz, J.T. 2002, ApJ 578, L161

\bibitem[1981]{Serio}
Serio, S., Peres, G., Vaiana, G.S. \& Rosner, R. 1981, ApJ 243, 288

\bibitem[1962]{Spitzer} Spitzer L. 1962, Physics of ionized gases.  Wiley, New
York

\bibitem[1999]{Teriaca} Teriaca, L., Banarjee, D. \& Doyle, J.G. 1999,  A\&A
349, 636

\bibitem[2002]{Testa} Testa, P., Peres, G., Reale, F. \& Orlando, S. 2002, ApJ
580, 1159

\bibitem[1988]{Thomas88} Thomas J.H. 1988, ApJ 333, 407

\bibitem[1990]{ThomasMontesinos90} Thomas J.H. \& Montesinos, B. 1990, ApJ
359, 550

\bibitem[1996]{Thomas96} Thomas J.H. in {\it Solar and Astrophysical MHD Flows}
ed. K.C. Tsinganos, Kluwer, Dordrecht, 39 (1996)

\bibitem[1982]{tsinganos1982} Tsinganos K.C. 1982, ApJ 252, 775

\bibitem[1993]{tsp93} Tsinganos K., Surlantzis G. \& Priest E.R. 1993, A\&A
275, 613

\bibitem[1993]{tts92} Tsinganos K., Trussoni, E. \& Sauty, C. 1992 in Schmelz,
J.T. \& Brown, J. eds, The Sun: A Laboratory for Astrophysicists, Kluwer,
Dordrecht, p. 349

\bibitem[1979]{Vesecky} Vesecky, J.F., Antiochos, S.K. \& Underwood, J.H. 1979,
ApJ 233, 987

\bibitem[1998]{VT98}
Vlahakis, N. \& Tsinganos, K. 1998, MNRAS 298, 777

\bibitem[1995]{WalshBellHood1995} Walsh, R.W., Bell, G.E. \& Hood, A.W. 1995,
Solar Phys. 161, 83

\bibitem[1996]{WalshBellHood1996} Walsh, R.W., Bell, G.E. \& Hood, A.W. 1996,
Solar Phys. 169, 33

\bibitem[2000]{WalshGaltier} Walsh, R.W. \& Galtier, S. 2000, Solar Phys. 197,
57

\bibitem[2002]{Winebarger} Winebarger, A.R., Warren, H.P. \& Mariska, J.T.
2002, ApJ in press

\bibitem[1981]{WraggPriest81} Wragg, M.A. \& Priest, E.R. 1981, Solar Phys.
70, 293

\bibitem[1982]{WraggPriest82} Wragg, M.A. \& Priest, E.R. 1982, Solar Phys.
80, 309

\end{thebibliography}
\end{document}